\documentclass[useAMS,usenatbib]{mn2e}

\usepackage{times}
\usepackage{amsmath,amssymb}
\usepackage{graphicx}
\usepackage{comment}

\newcommand{\apj}{ApJ}
\newcommand{\apjs}{ApJS}
\newcommand{\aj}{AJ}
\newcommand{\aap}{A\&A}

\newcommand{\mnras}{MNRAS}
\newcommand{\pasp}{PASP}
\newcommand{\icarus}{Icarus}
\newcommand{\nat}{Nature}

\newcommand{\araa}{ARA\&A}

\title[Gaia Planet Detection Potential around M Dwarfs]
{Astrometric Detection of Giant Planets Around Nearby M Dwarfs: The Gaia Potential}
\author[Sozzetti et al.]{A. Sozzetti$^1$, P. Giacobbe$^{1,2}$, M. G. Lattanzi$^1$, G. Micela$^3$, R. Morbidelli$^1$, and G. Tinetti$^4$\\
$^{1}$INAF - Osservatorio Astrofisico di Torino, Via Osservatorio 20, I-10025 Pino Torinese, Italy \\
$^{2}$Dept. of Physics, University of Trieste, Via Tiepolo 11, I-34143 Trieste, Italy \\
$^{3}$INAF-Osservatorio Astronomico di Palermo, P.za Parlamento 1, I-90134 Palermo, Italy \\
$^{4}$Department of Physics and Astronomy, University College London, Gower Street, London WC1E 6BT, UK}
\begin{document}

\date{Accepted ????. Received ???}

\pagerange{\pageref{firstpage}--\pageref{lastpage}} \pubyear{2011}

\maketitle

\label{firstpage}

\begin{abstract}
   Cool M dwarfs within a few tens of parsecs from the Sun are becoming the focus of dedicated  observational programs in the realm of exoplanet astrophysics. 
Gaia, in its all-sky survey  of $>10^9$ objects, will deliver precision astrometry for a magnitude-limited ($V=20$) sample of M dwarfs.
  We investigate some aspects of the synergy between the Gaia astrometric data on nearby M dwarfs and other ground-based and space-borne programs for planet detection 
and characterization. 
   We carry out numerical simulations to gauge the Gaia potential for precision astrometry of exoplanets orbiting a sample of known dM stars within $\sim30$ pc from the Sun. 
We express Gaia  detection thresholds as a function of system parameters and in view of the latest mission profile, including the most up-to-date astrometric 
error model. Our major findings are as follows: 
   (1) It will be possible to accurately determine orbits and masses for Jupiter-mass planets 
   with orbital periods in the range $0.2\lesssim P\lesssim 6.0$ yr and with an astrometric signal-to-noise ratio 
   $\varsigma/\sigma_\mathrm{AL}\gtrsim 10$. Given present-day estimates of the planet fraction $f_p$ around M dwarfs, $\approx 10^2$ 
   giant planets could be found by Gaia around the sample. Comprehensive screening by Gaia of the reservoir of  $\sim4\times10^5$ M dwarfs 
   within 100 pc could result in $\sim2600$ detections and as many as $\sim500$ accurate orbit determinations. The value of $f_p$ could then 
   be determined with an accuracy of 2\%, an improvement by over an order of magnitude with respect to the most precise values available to-date; 
   (2) in the same period range, inclination 
   angles corresponding to quasi-edge-on configurations will be determined with enough precision (a few percent) so that it will be possible to identify 
   intermediate-separation planets which are potentially transiting within the errors. 
   Gaia could alert us of the existence of 10 such systems. More than 250 candidates could be identified assuming solutions compatible with 
   transit configurations within 10\% accuracy, although a large fraction of these ($\sim85\%$) could be false positives; 
   (3) for well-sampled orbits, the uncertainties on planetary ephemerides, 
   separation $\varrho$ and position angle $\vartheta$, will degrade at typical rates of  $\Delta\varrho < 1$ mas yr$^{-1}$ and $\Delta\vartheta < 2$ deg yr$^{-1}$, respectively. 
   These are over an order of magnitude smaller than the degradation levels attained by present-day ephemerides predictions based on mas-level precision HST/FGS astrometry; 
   (4) planetary phases will be measured with typical uncertainties $\Delta\lambda$ of several degrees, resulting  (under the assumption of purely scattering atmospheres) 
   in  phase-averaged errors on the phase function $\Delta\Phi(\lambda)\approx0.05$, and expected uncertainties in the determination of the emergent flux of 
   intermediate-separation ($0.3<a<2.0$ AU) giant planets of $\sim20\%$.
   Our results help to quantify the actual relevance of the Gaia astrometric observations of the large sample of nearby M dwarfs in a synergetic effort 
   to optimize the planning and interpretation of follow-up/characterization measurements of the discovered systems by means of transit survey programs, 
   and upcoming and planned ground-based as well as space-borne observatories for direct imaging (e.g., VLT/SPHERE, E-ELT/PCS) and simultaneous multi-wavelength 
   spectroscopy (e.g., EChO, JWST).

\end{abstract}


\section{Introduction}\label{Intro}

In the search for an answer to one of the most fundamental questions of Mankind (`Are we alone?'), the nearest stars, within a 
few tens of pc from the Sun, provide the most obvious target sample. The fast-developing, highly interdisciplinary field of 
extrasolar planets has recently witnessed an increase in dedicated experiments aimed at cooler, low-mass M dwarfs, 
in addition to those focused on stars more like our Sun.
There are several important reasons for such a change in perspective, which can be summarized under two main themes: 
a) the shift in theoretical paradigms in light of new observations, and b) the improved understanding of the observational opportunities 
for planet detection and characterization provided by these stars. 

First, the observational evidence gathered by 
ultra-high-precision space-borne photometric surveys (e.g., Kepler),  although still a matter of debate (Fressin et al. 2013), 
indicates that the frequency of close-in ($P<50$ d) low-mass planets, 
i.e. Neptunes and Super-Earths, is an increasing function of decreasing stellar mass (Howard et al. 2012). This result has recently 
been strengthened by the findings of ground-based radial-velocity (RV) programs carried out with state-of-the-art facilities 
(e.g., HARPS): Super-Earths with $M_p\leq 10$ $M_\oplus$ within the Habitable Zone\footnote{In its standard definition, the Habitable 
Zone corresponds to the range of distances from a given star for which water could be found in liquid form on a planetary surface 
(Kasting et al. 1993)} (HZ) of low-mass stars appear ubiquitous (Bonfils et al. 2013). Very recent analyses of Kepler data have only 
further corroborated this evidence (Dressing \& Charbonneau 2013; Kopparapu 2013).  The identification of a rocky, 
habitable planet is the essential prerequisite to its possible characterization as an actual life-bearing celestial object. It is thus 
clear why low-mass M dwarfs, seen for long as providers of inhospitable environments for life (Huang 1959; Dole 1964), are now being moved at the center of 
the stage in the exoplanets arena (Scalo et al. 2007; Tarter et al. 2007, and references therein). 

Second, the sample of the nearest ($d< 25-30$ pc), relatively bright ($J<9-10$) M dwarfs is amenable to combined studies with a wide array 
of observational techniques, which can be exploited to the best of their potential providing the opportunity to characterize the architecture 
of planetary systems across orders of magnitude in mass and orbital separations in a way that's not readily achievable for Solar analogs. 
For example, the possibility to reach detection of short-period transiting rocky planets from the ground with modest-size telescopes 
($30-50$ cm class) is guaranteed by the small radii of M dwarfs, leading to deep transits ($\Delta m\gtrsim 0.005$ mag) for the case of 
planets with $2\leq R_p\leq 4$ $R_\oplus$ (e.g., Charbonneau et al. 2009). In addition, as we have discussed above, the favorable mass 
ratios allow for detection of rocky, potentially habitable planets with the RV technique, thanks to RV signals with amplitudes 
(a few  m s$^{-1}$) that are readily detectable with the most precise instruments available to-date. 
Analogously, at intermediate separations ($\approx 1-4$ AU) high-precision astrometry becomes sensitive 
to planets in the mass range between Neptune and Jupiter (e.g., Casertano et al. 2008). Finally, the favorable planet-star contrast ratios 
provided by the low intrinsic luminosity of M dwarfs allows for improved detectability thresholds of giant planets at wide separations ($>5-10$ AU) 
with direct imaging techniques (e.g., Bowler et al. 2012). For the same reason,  atmospheric characterization (via occultation spectroscopy) 
of transiting close-in Super-Earths can be achieved for this sample (e.g., Tessenyi et al. 2012). 

ESA's Cornerstone mission Gaia, with a present-day launch scheduled for November 2013, will carry out a magnitude limited ($V\leq20$), 
all-sky astrometric survey (complemented by onboard photometric and partial spectroscopic information) that is bound to revolutionize 
our understanding of countless aspects of astronomy and astrophysics within our Milky Way, and beyond (e.g., Perryman et al. 2001). The global impact 
of Gaia micro-arcsecond-level ($\mu$as) astrometric measurements in the astrophysics of planetary systems has been addressed in the past 
(e.g., Lattanzi et al. 2000; Sozzetti et al. 2001, 2003; Casertano et al. 2008; Sozzetti 2011). However, those studies only provided general 
metrics for gauging detectability thresholds as a function of planetary properties (orbital elements, masses), using solar-like stars 
as the reference primaries. In addition, only brief mentions were made of the potentially huge levels of synergy between Gaia astrometry 
and other ongoing and planned exoplanet search and characterization programs. The approach adopted to carry out the analysis, 
particularly at the level of single- and multiple-planets orbital solutions, was still affected by some caveats and simplifying assumptions 
(e.g., only partial treatment or complete neglection of the problem of identifying adequate starting values for the non-linear fits). 
Finally, the Gaia astrometric performance, described in those works through a simple Gaussian single-measurement error model, has further evolved.  A more 
realistic error model, which takes into account e.g. the dependence on magnitude, ought to be utilized. 

 In this work, we revisit the topics of planet detection and characterization with Gaia relaxing some of the above assumptions, and 
focusing on the sample of nearby low-mass M dwarf stars for which Gaia, as one by-product of its all-sky survey, will deliver precision astrometry 
down to the $V=20$ magnitude limit. The main thrust of this paper is two-fold. 

First, we will gauge the Gaia potential for precision astrometry of exoplanets orbiting 
an actual sample of thousands of known dM stars within $\sim30$ pc from the Sun (L\'epine 2005). We will then express Gaia sensitivity 
thresholds as a function of system parameters and in view of the latest mission profile, including the most up-to-date astrometric error model. 
The analysis of the simulations results will also provide insight on the capability of high-precision astrometry 
to reconstruct the underlying orbital element distributions and occurrence rates of the planetary companions. These results will help 
in evaluating the expected Gaia recovery rate of actual planet populations around late-type stars. 

Second, we will investigate some elements of the synergy between the Gaia data on nearby M dwarfs and other ground-based and space-borne programs for planet 
detection and characterization, with a particular focus on: a) the potential for Gaia to precisely determine the orbital inclination, 
which might indicate the existence of transiting long-period planets; b) the ability of Gaia to accurately predict the ephemerides of 
(transiting and non-transiting) planets around M stars, and c) its potential to help in the precise determination of the emergent flux, 
for direct imaging and systematic spectroscopic characterization of their atmospheres with dedicated observatories from the ground and 
in space. 

Our paper is organized as follows. In \S~\ref{setup} we describe the adopted simulation setup, and in \S~\ref{tools} we present the 
statistical and numerical tools used to analyze the simulated datasets. \S~\ref{analisi} is devoted to the analysis of the simulation 
results. Finally, we summarize in \S~\ref{summ} our findings and provide concluding remarks.


\section{Simulation Scheme}\label{setup}

\begin{figure*}
\centering
$\begin{array}{cc}
\includegraphics[width=0.45\textwidth]{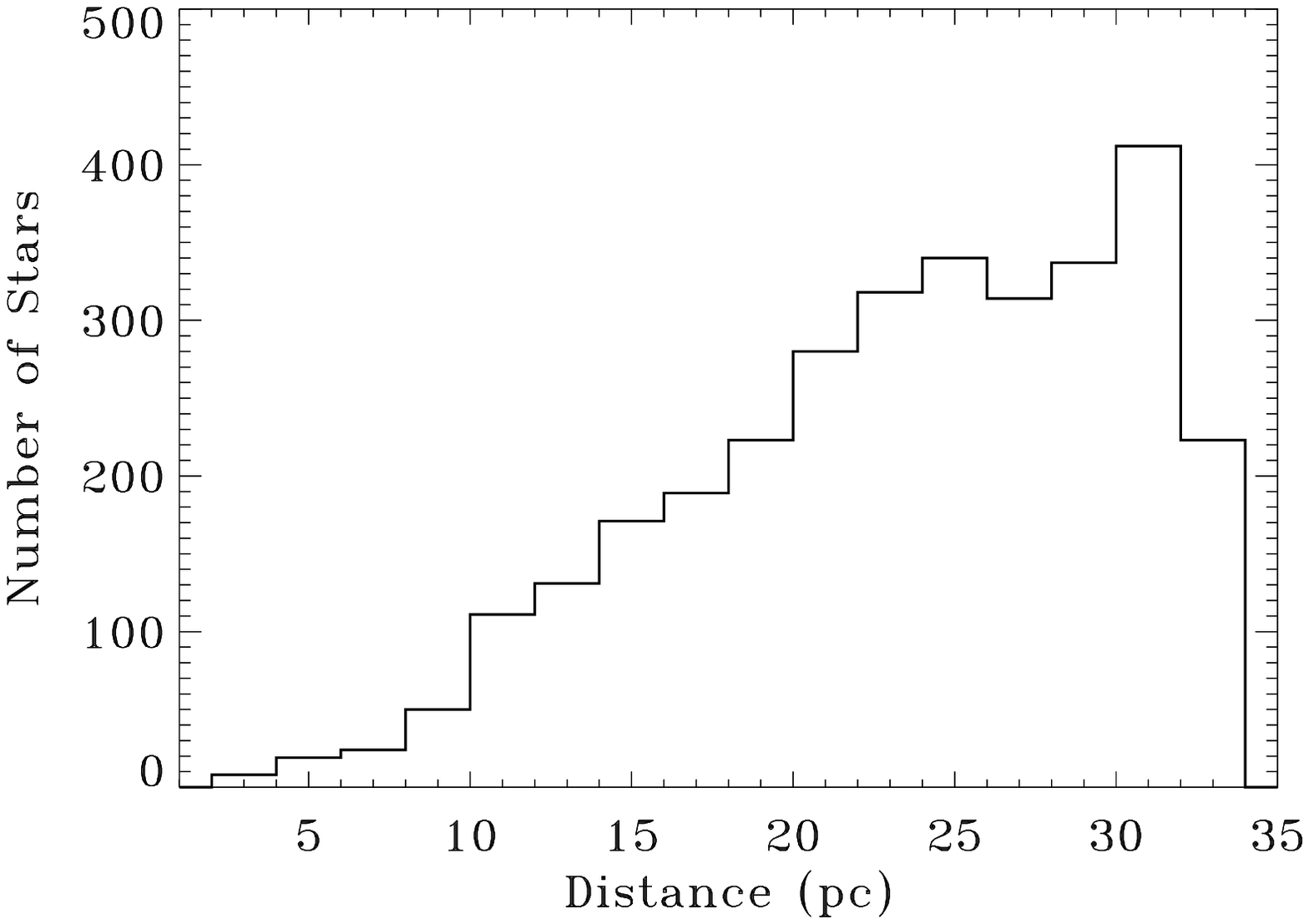} & 
\includegraphics[width=0.45\textwidth]{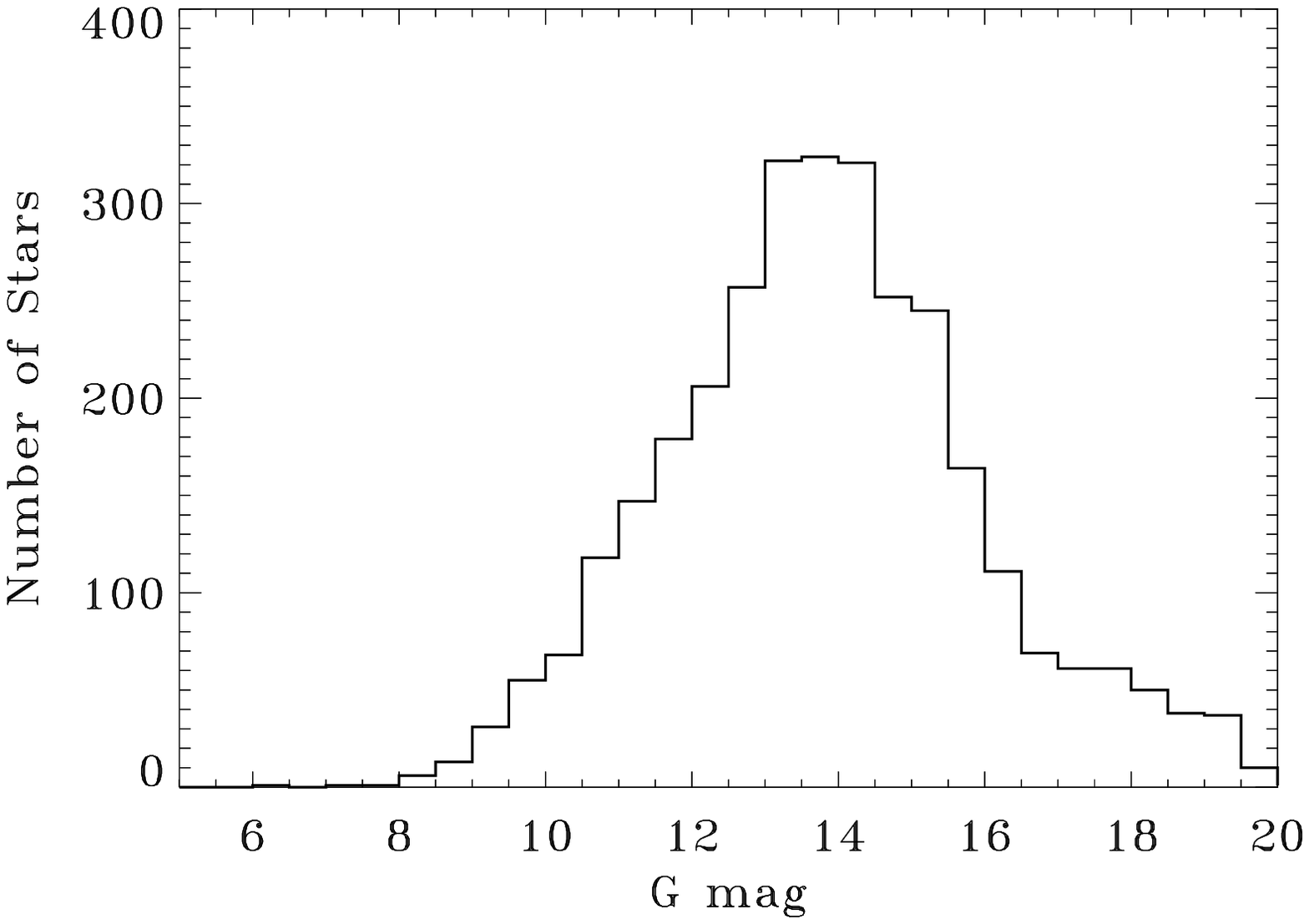} \\
\includegraphics[width=0.45\textwidth]{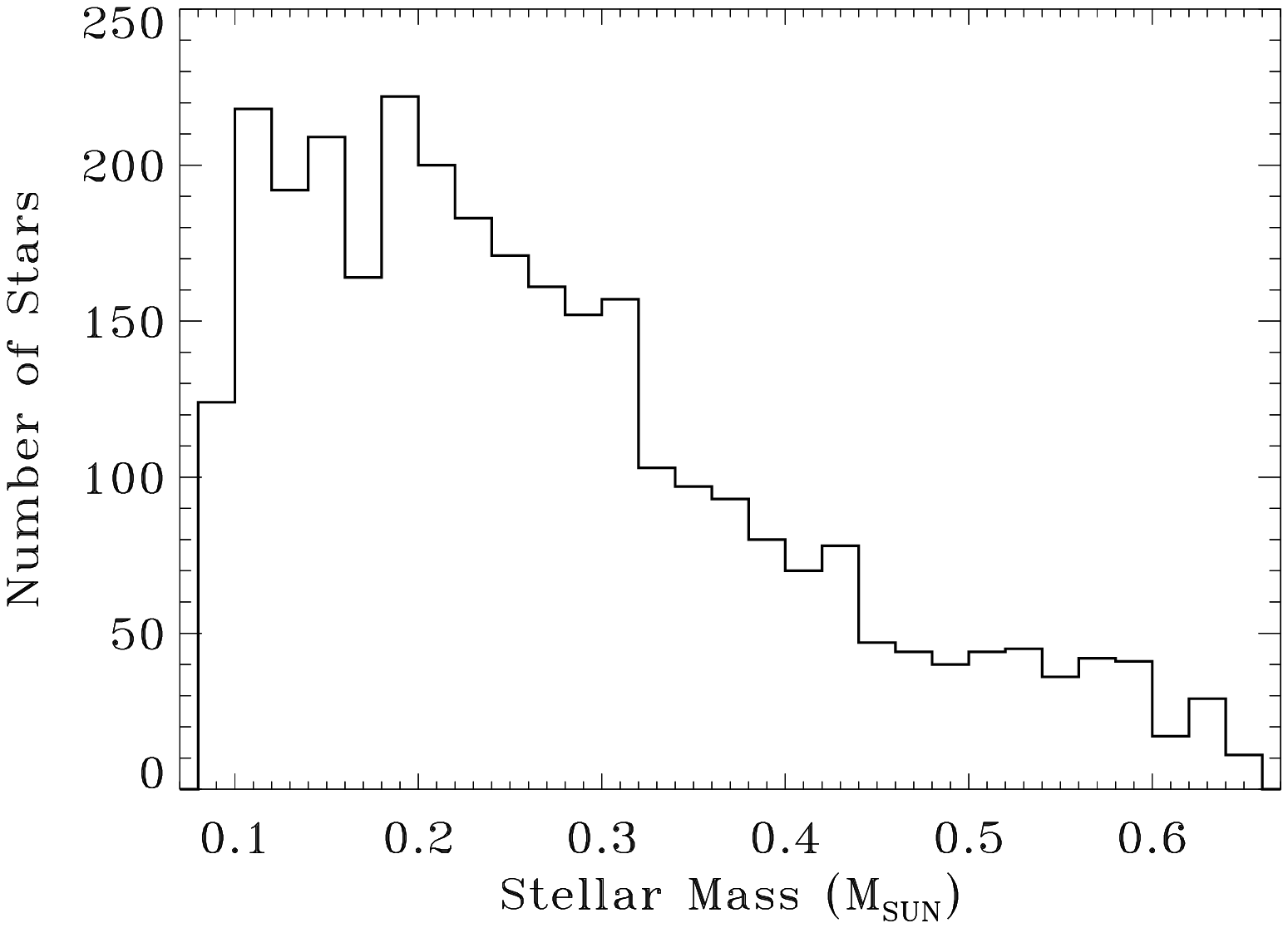} & 
\includegraphics[width=0.45\textwidth]{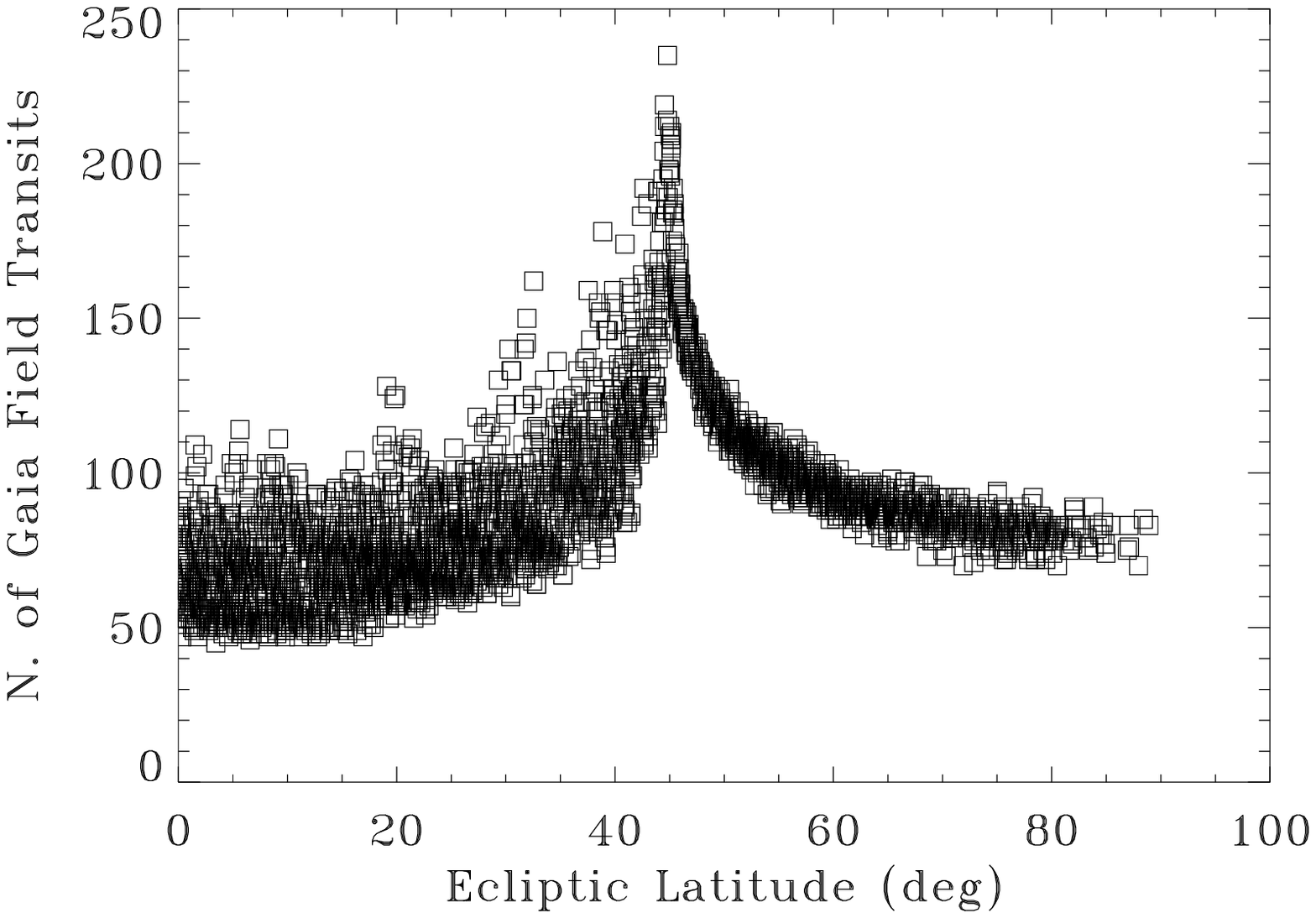} \\
\end{array} $
 \caption{Top left: distance distribution of the  LSPM M dwarf sub-sample. Top right: The corresponding magnitude distribution in Gaia's $G$ band. 
 Bottom left: the derived stellar mass distribution. Bottom right: number of individual Gaia field transits for the sample.}
\label{fig1}
\end{figure*}

The simulation of Gaia observations follows closely the observational scenario described in Casertano et al. (2008). We refer the reader 
to that source for details. Here we describe and discuss the changes/upgrades made to that setup. 

\begin{figure}
\centering
\includegraphics[width=0.50\textwidth]{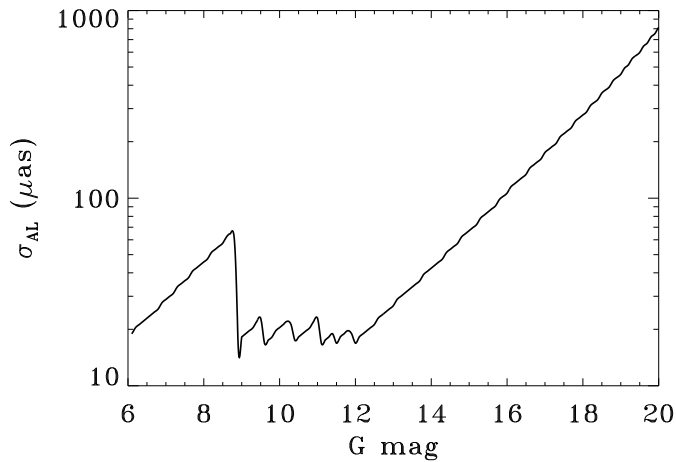}
 \caption{Gaia single-transit astrometric uncertainty as a function of $G$-band magnitude, including the effects 
 of the adopted gate scheme.}
\label{fig2}
\end{figure}

\begin{itemize}
\item[1)] In the representation of the Gaia satellite, 
the latest nominal Gaia scanning law was utilized, with the two fields of view separated by a basic angle of 106.5 deg, 
with a spin rate of 60 arcsec s$^{-1}$, a solar aspect angle  between the direction to the Sun and the satellite's spin axis $\xi=45$ deg, 
and a precessional period of the spin axis of 63 days. Details on the scanning geometry of the Gaia satellite can be found in, e.g., Lindegren (2010) 
and Lindegren et al. (2012). The nominal mission duration ($T=5$ yr) was adopted. 

\item[2)] The actual list of targets encompasses 3150 low-mass stars (in the approximate range $0.09-0.6$ $M_\odot$) within 33 pc from the Sun  (L\'epine 2005) 
from the proper-motion limited LSPM-North Catalog  (L\'epine \& Shara 2005). For convenience we are referring to this sample collectively as M dwarfs, 
even though some of them have estimated masses more compatible with those of late K dwarfs. 
This subset of the LSPM catalog (dubbed LSPM sub-sample hereafter) is not complete within the identified volume limit, with as much as 32\% of 
stars missing out to 33 pc (L\'epine 2005). However, the choice of this catalog over, for example, the more recent L\'epine \& Gaidos (2011) all-sky catalog was 
driven by our interest to choose a  volume-confined sample (so that distance effects in the detectability of astrometric signals can be 
more simply taken into account). Using visual and infrared magnitudes available for the sample, we utilized the color-magnitude 
conversion formulae of Jordi et al. (2010) to obtain $G$-band magnitudes in the Gaia broad-band photometric system. The  LSPM sub-sample results 
to have an average $G\simeq14.0$ mag. The Delfosse et al. (2000) mass-luminosity relations for low-mass stars were then utilized to 
obtain mass estimates for all our targets. The  LSPM sub-sample results to have an average $M_\star\simeq0.30$ M$_\odot$. 
In the four panels of Fig.~\ref{fig1} we show the distributions in $G$ mag, distance $d$
\footnote{Where available, Hipparcos parallaxes are used, photometric distance estimates are otherwise utilized using the L\'epine (2005) values.} 
mass $M_\star$, and number of Gaia field transits (individual field-of-view crossings) as a function of ecliptic latitude $\beta$ 
for our  LSPM sub-sample. The dependence of the number of Gaia measurements with $\beta$, with the maximum in correspondence of $\beta=\xi$, 
is a result of the adopted scanning law (see e.g. Lindegren et al. 2012 for details). 

\item[3)] The $G$-band magnitudes were then used to derive the resulting along-scan single-measurement uncertainty $\sigma_\mathrm{AL}$ 
using an up-to-date magnitude-dependent error model (e.g., Lindegren 2010). The presently envisioned gate scheme to avoid saturation on bright stars 
(affecting some 20\% of $G<12$ mag M dwarfs) was included in the calculation (see Fig.~\ref{fig2}), but provisions were not made for 
the representation of charge-transfer-inefficiency effects in the measurements. Single-measurement errors are typically $\sigma_\mathrm{AL}\sim100$ $\mu$as. 
Given the typical magnitude of the Gaia positional uncertainties involved in the simulations, the astrometric 'jitter' induced by 
spot distributions on the stellar surface of active M dwarfs 
(e.g., Sozzetti 2005; Eriksson \& Lindegren 2007; Makarov et al. 2009; Barnes et al. 2011) 
was considered to be negligible, and  therefore not included in the error model.

\item[4)] The generation of planetary systems proceeded as follows. One planet was generated around each star (assumed not to be orbited by a stellar companion), 
with mass $M_p = 1 M_J$, orbital periods were uniformly distributed in the range $0.01\leq P\leq 15$ yr and eccentricities were uniformly distributed 
in the range $0.0\leq e\leq0.6$). The orbital semi-major axis $a_p$ was determined using Kepler's thid law. 
All other orbital elements (inclination $i$, argument of pericenter 
$\omega$, ascending node $\Omega$, and epoch of pericenter passage $\tau$) were uniformly distributed within their respective ranges 
(for the inclination $\cos i$ was uniformly distributed). 
The resulting astrometric signature induced on the primary was calculated using the standard formula corresponding to the 
semi-major axis of the orbit of the primary around the barycenter of the system scaled by the distance to the observer: 
$\varsigma=(M_p/M_\star)\times(a_p/d)$. With $a_p$ in AU, $d$ in pc, and $M_p$ and $M_\star$ in M$_\odot$, then $\varsigma$ is evaluated in arcsec. 
Note that $\varsigma$ corresponds to the true perturbation size only in the case of circular orbits. It is in general only an upper limit 
to the actual magnitude of the measured perturbation when projection and eccentricity effects are taken into account (e.g., Sozzetti et al. 2003; 
Reffert \& Quirrenbach 2011).

\end{itemize}

\section{Statistical and Numerical Analysis Tools}\label{tools}

The tools utilized in the analysis of the simulated Gaia astrometric data have already been described elsewhere (Casertano et al. 2008). 
We briefly recall here their main features. First, statistically robust deviations from a single-star model, indicating the presence in the 
observations residuals of the perturbation due to a companion with a given level of confidence, are identified through the application of 
a $\chi^2$-test or $F$-test (low probabilities of $P(\chi^2)$ or $P(F)$ signifying likely planet, and unlikely false positive). 
Then, orbital fits to the data are carried out, using a Markov Chain Monte Carlo (MCMC)-driven global search approach to the identification of good starting guesses for the 
orbit fitting procedure that combines a period search with a local minimization algorithm (Levenberg-Marquardt). Details on the overall algorithm 
performance as applied to large datasets of synthetic Gaia observations produced within the context of the Gaia Data Processing and Analysis Consortium (DPAC)
\footnote{http://www.rssd.esa.int/gaia/dpac} will be published elsewhere. As described in Casertano et al. (2008), 
the resulting Gaia observable, the one-dimensional coordinate $\psi$ in the along-scan direction of the instantaneous great circle followed by Gaia at
that instant, will then be modeled as $\psi(\alpha, \delta, \mu_\alpha, \mu_\delta, \pi, A, B, F, G, P, e, \tau)$, where the five standard astrometric 
parameters correspond to the actual positions $(\alpha, \delta)$, proper motion components $(\mu_\alpha, \mu_\delta)$, and parallax ($\pi$) of 
each target M dwarf as provided in L\'epine (2005), while $A$, $B$, $F$, and $G$ are four of the six Thiele-Innes elements (Green 1985). Planetary masses 
are derived from the best-fit orbital elements assuming perfect knowledge of the stellar primary mass and utilizing the approximation 
of the mass-function formula (valid in the limit $M_p\ll M_\star$):

\begin{equation}
 M_p\simeq \left(\frac{a_\star^3}{\pi^3}\frac{M_\star^2}{P^2}\right)^{1/3},
\end{equation}
with $M_\star$ in solar-mass units, $P$ in years, $\pi$ and $a_\star$ (the semimajor axis of the orbit of the central star around the barycenter) 
both expressed in arcseconds. 

\section{Results}\label{analisi}

\subsection{Detection Probabilities}\label{detect}

\begin{figure*}
\centering
\includegraphics[width=0.85\textwidth]{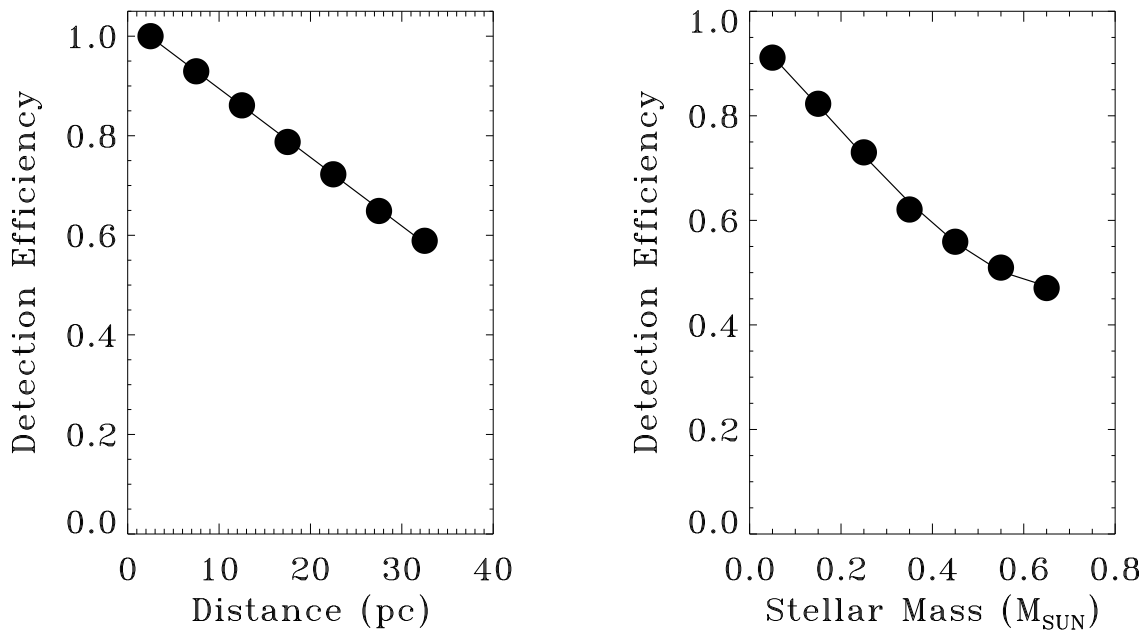} 
 \caption{Left: Gaia giant planet detection efficiency around a sample of 3150 M dwarfs from L\'epine (2005) as a function of distance 
 from the Sun (data are binned in 5 pc bins). 
 Right: The same quantity expressed as a function of primary mass (data are binned in 0.1 M$_\odot$ bins).
 }
\label{fig3}
\end{figure*}

The probability  of a planet being detected is obtained, following Lattanzi et al. (2000) and Casertano et al. (2008), via a $\chi^2$-test of 
the null hypothesis that a star is single. By setting a probability threshold $P(\chi^2)\leq0.001$  (i.e., a confidence level of 99.9\%), we 
find that 2704 stars (85\% of the sample) are identified as variable. The detected systems have an `astrometric signal-to-noise ratio' 
$\varsigma/\sigma_\mathrm{AL}\gtrsim3$. 
The ratio of the perturbation size to the single-measurement accuracy had already been shown to be of use by Sozzetti et al. (2002) to study the main trends 
in astrometric planet detection probabilities and the limits for accurate orbit determination.  For comparison, both Sozzetti et al. (2002) and Casertano et al. (2008) 
found that $P(\chi^2)\leq0.05$ (i.e., a confidence level of 95\%) would enable detection of astrometric signatures with $\varsigma/\sigma_\mathrm{AL}\gtrsim2$.
As already discussed in Lattanzi et al. (2000) and Casertano et al. (2008), 
the standard experiment to gauge the false alarm rate in the presence of pure white noise due to the adopted statistical threshold for detection gave the 
expected results (e.g., 1\% of false positives for a $\chi^2$-test with a confidence level of 99\%). 

The behaviour of detection probabilities (at the $\varsigma/\sigma_\mathrm{AL}\geq3$ level) with stellar mass and distance for the  LSPM sub-sample at hand is shown in Figure~\ref{fig3}. The left panel 
shows the trend of detection efficiency with distance from the Sun marginalizing over all primary mass values, while the right panel presents the trend with stellar mass marginalizing 
over distance.  We find that detection efficiency decreases linearly with distance and for  the LSPM sub-sample detection rates of $\approx60$\% are still achieved at the distance limit of $\sim33$ pc. 
At the high-mass end, this number falls below 50\%. The lowest-mass stars ($M_\star<0.1$ M$_\odot$) have the highest detection rate, which is not surprising given that in  the LSPM sub-sample, 
while being the faintest, they are the closest neighbors to the Sun. 

\begin{figure}
\centering
\includegraphics[width=0.49\textwidth]{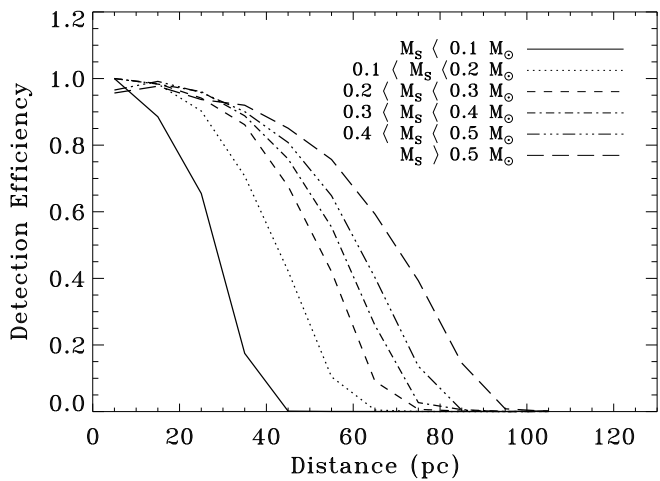} 
 \caption{The dependence of giant planet detection efficiency with Gaia astrometry as a function of distance and primary mass (lines of different styles) 
 based on M dwarfs starcounts out to 100 pc from the Sun derived using the Besancon stellar population synthesis model. 
 }
\label{fig4}
\end{figure}

The same analysis can be made more robust in a statistical sense by using stellar population models to estimate the number of M dwarfs that Gaia will observe and estimate 
the rate of detection in the entire  sample accessible to Gaia. We derived starcounts for M0-M9 dwarfs with $G<20$ and within 100 pc from the Sun 
using the Besancon stellar population synthesis model (Robin et al. 2003), 
taking advantage of a recent correction (March 2013) applied on the spectral types and luminosity function of late-M dwarfs (see http://http://model.obs-besancon.fr/). 
The total sample amounts to $\sim415,000$ stars, with effective temperatures T$_\mathrm{eff}$ in the range between 2500 K and 4000 K, 
masses in the range between 0.08 M$_\odot$ and 0.63 M$_\odot$, absolute visual magnitudes $M_v$ in the range between 8.5 and 19.0, and colors $V-I$ in the range 
between 1.55 and 4.65. Figure~\ref{fig4} shows how detection efficiency (when $\varsigma/\sigma_\mathrm{AL}\gtrsim3$) varies with distance and primary mass for the 
 simulated sample from the Besancon galaxy model (BGM sample for short henceforth). 
The distance horizon for 90\% completeness around the lowest-mass bin is $\sim15$ pc, and $\sim40$ pc for M0 dwarfs. Detectability drops below 50\% for M8-M9 dwarfs 
at $\sim30$ pc, while for stars with $M_\star>0.5$ M$_\odot$ the same detection efficiency levels extend out to $\sim80$ pc. The above considerations can be turned into an actual number of 
detected giant planets across all spectral sub-types of M dwarfs, as we will see later on in \S~\ref{yield}. 

\begin{figure}
\centering
\includegraphics[width=0.45\textwidth]{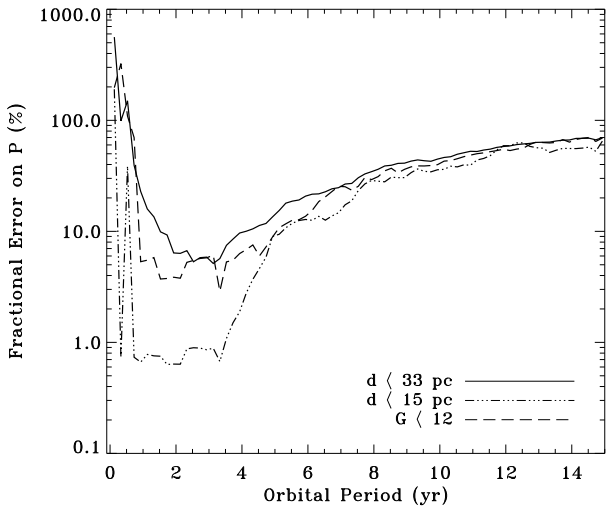} 
 \caption{Fractional error on the orbital period $P$ as a function of $P$ itself. Solid line: the full  LSPM sub-sample. Dashed-dotted line: 
 stars $d<15$ pc. Dashed line: stars with $G<12$ mag.
 }
\label{fig5}
\end{figure}

\subsection{Orbit Determination}\label{orbit}

\begin{figure}
\centering
$\begin{array}{c}
\includegraphics[width=0.45\textwidth]{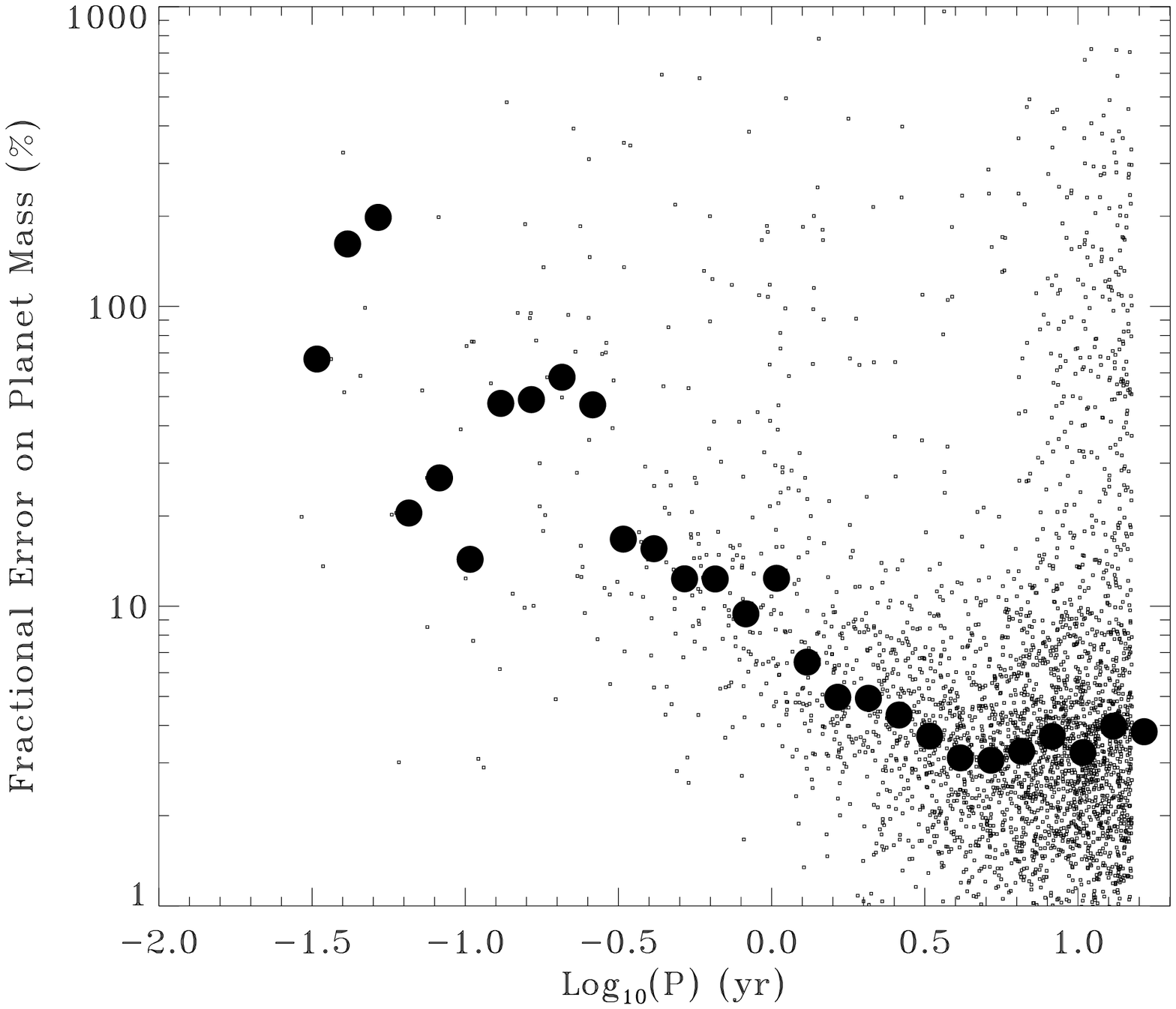}\\ 
\includegraphics[width=0.45\textwidth]{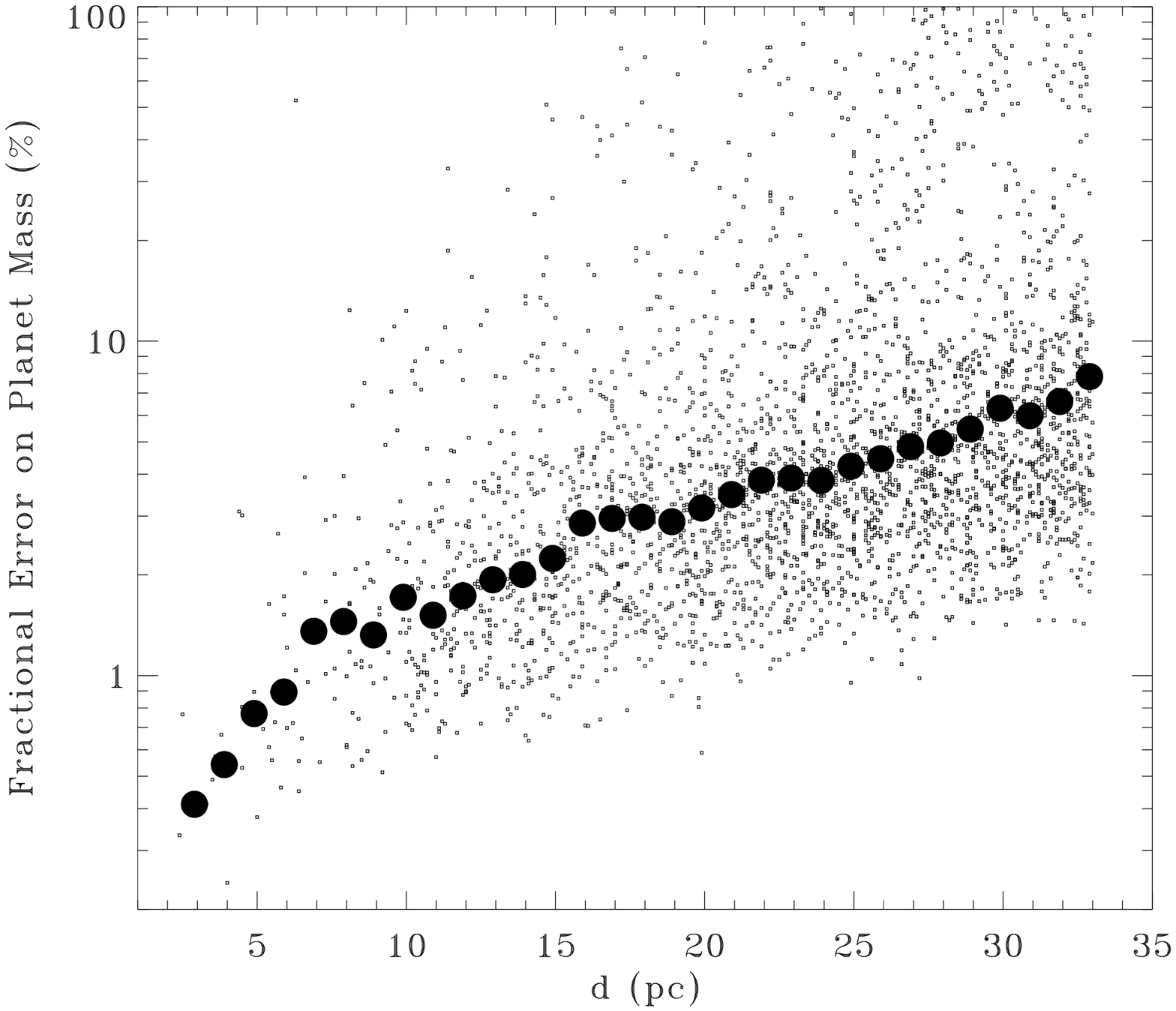} \\
\end{array} $
 \caption{Top: Fractional error on planetary mass as a function of orbital period. Bottom: the same, 
 but as a function of distance from the Sun. In both cases, the large black dots represent the 
 median value in a period or distance bin, respectively.
}
\label{fig6}
\end{figure}

Casertano et al. (2008) presented detailed analyses of the sensitivity of Gaia astrometry to giant planets around solar-type stars. 
We discuss here the quality of orbit reconstruction for the case of Jupiter-mass companions orbiting the  LSPM sub-sample, 
using as proxies the fitted orbital periods and the derived planetary masses, and the precision with which these parameters 
are retrieved as a result of the orbit fitting procedure, expressed in terms of the fractional error (e.g., $(P_\mathrm{fitted}-P_\mathrm{true})/P_\mathrm{true}$). 
We refer to an accurate determination of a given parameter when its fractional error is less or equal to 10\%. The quality of orbit 
reconstruction will also be parameterized in terms of $\varsigma/\sigma_\mathrm{AL}$ (e.g., Sozzetti et al. 2002). 

We show in Fig.~\ref{fig5} the variation of the fractional error on the orbital period as a function of the true simulated value of $P$. 
As expected, some of the main features of this behaviour already described in Casertano et al. (2008) are recovered. For example, Fig.~\ref{fig5} 
highlights how Gaia sensitivity decreases significantly both for periods exceeding the mission duration as well as for short-period orbits which are 
under-sampled (as a direct effect of the scanning law) and translate in very low astrometric signals. 
On the other hand, well-sampled ($P<T$) orbital periods can be determined with uncertainties 
of $<10\%$ around the nearest sample ($d<15$ pc, approximately 450 targets). In the same range of periods, the precision improves if 
a magnitude cut-off ($G\leq12$, approximately 600 targets) is made, but not to a very significant extent. 
Bright objects are in fact somewhat affected (in terms of planet detectability and 
quality of orbit reconstruction) by the presently envisioned gate scheme to avoid saturation on bright stars (see Fig.~\ref{fig2}). 
Instead, at least for the  LSPM sub-sample under investigation, the nearest stars ($d<15$ pc) appear to provide the most significant improvement in precision 
in orbital period determination.  The resulting astrometric signatures are typically large enough to allow for good-accuracy orbit reconstruction 
even for relatively faint objects, for which the per-measurement precision is significantly degraded. 

\begin{figure}
\centering
\includegraphics[width=0.50\textwidth]{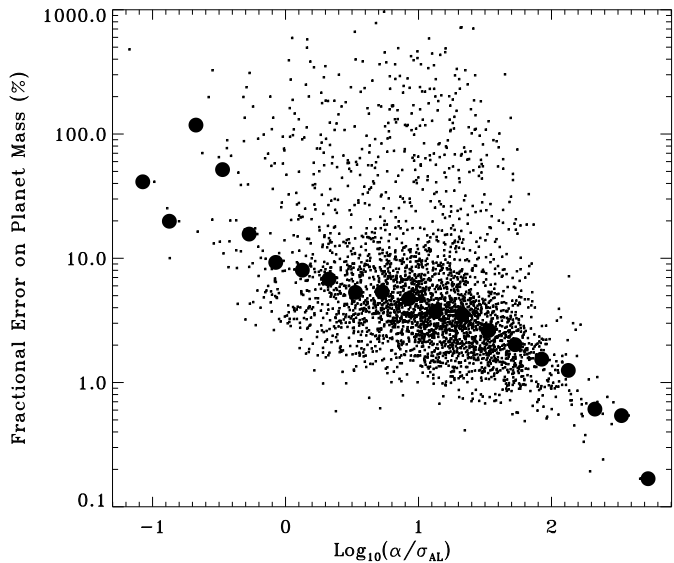} 
 \caption{Fractional error on planetary mass as a function of astrometric signal-to-noise ratio  (all stars included). Black dots represent the median 
 values in each $\varsigma/\sigma_\mathrm{AL}$ bin.
}
\label{fig7}
\end{figure}

The planetary mass as derived using the mass function approximation will be affected by the uncertainty on $P$, $\pi$, and $a_\star$ as obtained from the 
fitting procedure.\footnote{We assume here that $M_\star$ is perfectly known. While uncertainties in stellar mass at the bottom of the 
main sequence can easily be on the order of 10-20\% (e.g., Boyajian et al. 2012 and references therein), the uncertainties in the model parameters 
from orbital fits are the dominant source of error when deriving the companion mass in this analysis.}
Its median for the whole  LSPM sub-sample is 1.19 M$_J$, which reduces to $0.99$ M$_J$ for the sample within 20 pc from the Sun. The two panels 
of Fig.~\ref{fig6} show how $P$ and $d$ affect the uncertainty on M$_p$. In particular, planets orbiting stars within $\sim20$ pc have 
their masses measured with typical precision of $10\%$, or better, while short- and long-period orbits allow for reduced precision in the 
derivation of M$_p$, as expected. For $P>T$, planetary masses are systematically overestimated as a result of the systematic under-estimation of $P$, 
an effect already shown and discussed in detail by Casertano et al. (2008). 

One representation of the dependence of the quality of orbit reconstruction on $\varsigma/\sigma_\mathrm{AL}$ is visualized in Fig.~\ref{fig7}. 
From the plot, one sees for example that $\varsigma/\sigma_\mathrm{AL}\simeq 10$ allows to measure $M_p$ to $\sim 10\%$ accuracy, 
while fractional errors of a few percent, or better, require $\varsigma/\sigma_\mathrm{AL}\gtrsim 30-40$. 
These findings are in line with those obtained by Sozzetti et al. (2002).  In Fig.~\ref{fig7}, the results of orbital fits for the full  LSPM sub-sample are plotted, 
including increasingly more marginal detections ($\varsigma/\sigma_\mathrm{AL} < 3$). As expected, for some systems (with $1\lesssim\varsigma/\sigma_\mathrm{AL}\lesssim3$) 
it might still be possible to obtain reasonably accurate solutions and mass estimates (for example in cases of particularly favorable orbital sampling). 
However, in general, attempting to fit orbital signals with amplitudes too close to, or even below, the single-measurement precision will translate in very poor 
quality results.

Table~\ref{tab1} reports, for illustrative purposes, the relevant information for 8 giant planetary companions to nearby M dwarfs detected by Doppler surveys (as 
obtained from http://exoplanet.eu). From the the minimum $\varsigma_\mathrm{min}/\sigma_\mathrm{AL}$
\footnote{We recall that for Doppler-detected planets only the lower limit $M_p\sin i$ to their mass can be determined.} values, one infers that, 
given the general systems characteristics (orbit, masses, distance), half of the sample of known Doppler-detected 
planets around M dwarfs considered here  (GJ 832b, GJ 849b, GJ 179b, GJ 317b) would have accurately determined orbital parameters and masses based on Gaia astrometry 
alone, while one of the planets,  HIP 57050b, would be essentially undetectable. Note, however, that this situation corresponds to the worst-case scenario, 
as the actual masses will typically be larger than the minimum value reported in Table~\ref{tab1}. The two short-period, resonant giant planets orbiting GJ 876 
would both be detectable by Gaia, but they are not included in this sample as our study does not cover the problem of multiple-planet detection and characterization 
around low-mass stars. 
Other M dwarf planets have not been considered as their inferred astrometric signals (due to a combination of low masses, short periods, and large distances) 
would be essentially undetectable in Gaia astrometry. 

\subsection{Expected Planet Yield}\label{yield}

 It is worthwhile providing a reference figure of merit on the number of giant planets we can expect Gaia to detect in a given interval of 
orbital separations, as a way of gauging, in a preliminary fashion, the ability of the survey to reconstruct the underlying orbital elements distributions 
and occurrence rates in the low-mass star regime. In two recent works, Johnson et al. (2010b) and Bonfils et al. (2013) have provided updated estimates of 
the fraction of M dwarfs (no distinction in the stellar sub-types given the small-number statistics involved) hosting giant planets within approximately 3 AU. 
Starting with a northern hemisphere sample observed with HIRES and a southern hemisphere sample observed with HARPS, with different minimum-mass 
sensitivity thresholds but comparable time baselines, they reach similar conclusions: short-period ($P<100$ days) giants ($M_p\sin i>0.3$ $M_J$) 
are quite rare around M dwarfs ($f_p<1\%$). At wider separations (roughly, $a<3$ AU), giants orbiting M dwarfs appear to be more frequent: Johnson et al. (2010b) report 
$f_p\simeq3\pm1\%$ (corrected for metallicity effects), while Bonfils et al. (2013) obtain $f_p\simeq4^{+5}_{-1}\%$, two estimates which appear consistent with each other, 
within the error-bars. Note that these values of $f_p$ are somewhat higher than those quoted in previous works.  For example, Endl et al. (2006) derive an upper limit 
(at the $1-\sigma$ confidence level) of $f_p\simeq1.3\%$ for giant planets within 1 AU of low-mass stars, while Cumming et al. (2008) infer $f_p\simeq2\%$ 
for M dwarfs orbited by gas giants within $\sim3$ AU. It is furthermore worth pointing out how giant planet occurrence rates at intermediate 
separations ($\sim3$ AU) around M-dwarf hosts from microlensing surveys (e.g., Gould et al. 2010) appear reasonably in agreement with the above results. 
 Other recent microlensing and high-contrast imaging studies encompassing the range of orbital separations $a<20$ AU for gas giants provide roughly 
consistent numbers ($f_p\simeq9^{+3}_{-5}\%$ from Cassan et al. (2012) and $f_p\simeq6.5\pm3.0\%$ from Montet et al. (2013), respectively).

We can estimate the number of giant planets at a given distance d (in pc) whose astrometric signal could be detected by Gaia using the $\varsigma/\sigma_\mathrm{AL}\gtrsim3$ 
detectability threshold (\S~\ref{detect}), starcounts $N_\star$ for M dwarfs computed within a sphere of radius d centered on the Sun, and the 
value of $f_p=3\%$ derived by Johnson et al. (2010b).
We can then infer that around the nearby ($d<33$ pc)  LSPM sub-sample under consideration here, Gaia should be able to detect on the order of 100 giant planets 
within 3 AU, an order-of-magnitude increase with respect to the present-day yield from Doppler surveys. Given the average stellar mass of the sample, 
the corresponding orbital period limit would be $P\lesssim6$ yr, thus this planet sample would also be one for which Gaia could deliver good-quality orbital 
solutions (at least for the brighter/nearer systems). Extrapolating to the  BGM sample of M dwarfs within 100 pc from the Sun from the Besancon Galaxy model presented 
in \S~\ref{detect}, 
one then infers a total giant planet yield for Gaia of $\approx2600$ new detections. Accurate orbit reconstruction will be possible for $\approx20\%$ of the 
detections (corresponding to systems with $\varsigma/\sigma_\mathrm{AL}\gtrsim 10$), i.e $\sim500$ giant planetary companions orbiting M dwarfs within $\sim50$ pc. 
Naturally, this estimate does not take into account any possible dependence of occurrence rates and orbital elements and mass distributions on spectral sub-type 
(for example, for M5 or later dwarfs, $f_p$ is presently severely unconstrained by observations), 
nor do we attempt at extrapolating at wider orbital separations. 

Finally, while theoretical arguments based on the core-accretion model of giant planet formation 
(e.g., Laughlin et al. 2004; Ida \& Lin 2005; Alibert et al. 2011) clearly  predict the existence of a trend of decreasing $f_p$ with decreasing 
$M_\star$, as observed (Johnson et al. 2010b), the predicted planet fractions in a given stellar mass range do not necessarily 
agree with the observations. For example, Kennedy \& Kenyon (2008) predict $f_p\sim1\%$ within $\sim2-3$ AU of $M_\star<0.5$ M$_\odot$ M dwarfs, 
a value somewhat lower than the observed fraction.  Any discrepancy could point to either insufficient depth in the analysis of the observational data 
or to the necessity to further the theoretical understanding of planet formation processes in the low-mass star regime. However, at present any attempt to study 
fine structure details in the comparison between theory and observations is severely hampered by small-number statistics. In this respect, Gaia high-precision astrometry 
of thousands of nearby M dwarfs will likely help to shed light into the matter, as this unbiased sample screened for giant planets by Gaia will contribute to significantly reduce the uncertainties on the 
occurrence rate estimates.  For example, based on the above simulation results we can infer how precisely a value of $f_p=3\%$ for Jupiter-mass companions 
within 3 AU around M0-M9 stars could be determined by using the number of detections ($n\simeq2600$) and the number of stars for which an astrometric detection was possible ($N\simeq9\times10^4$), 
the latter derived based on the detection efficiency estimates presented in \S~\ref{detect} using the Besancon galaxy model. Using this non-parametric description (see e.g., 
Cumming et al. 2008) and by adopting the standard Poisson uncertainty limits $(1/n+1/N)^{1/2}$ (e.g., Burgasser et al. 2003), 
we then obtain $f_p=3.00\pm0.06$, an improvement by a factor $\sim15$ with respect to, e.g., the Johnson et al. (2010b) estimates. 

\begin{table*}
\centering
\caption{Summary of relevant data on known Doppler-detected giant planets orbiting $M_\star<0.6$ M$_\odot$ dwarfs in the solar neighborhood.}
\begin{tabular}{lcccccccccc}
 Name & $G$ & $\varsigma_\mathrm{min}$ & $\sigma_\mathrm{AL}$ & $\varsigma_\mathrm{min}/\sigma_\mathrm{AL}$ & $P$ & $e$ & $d$ & $M_\star$ & $M_p\sin i$ & Discovery\\
 & (mag)& ($\mu$as) & ($\mu$as) &  & $(yr)$ & & (pc) & ($M_\odot$) & ($M_J$) & (Refs)\\  
\hline
 GJ 832b   &   8.7 &   978.9  &   64.4  &   15.2  &    9.3 &  0.12 &   4.9  &    0.45 &     0.64 & Bailey et al. (2009)\\
 Gl 649b   &   9.7 &    66.7  &   17.5  &    3.8  &    1.6 &  0.30 &  10.3  &    0.54 &     0.33 & Johnson et al. (2010a)\\
 GJ 433c   &   9.8 &   116.1  &   18.1  &    6.4  &   10.1 &  0.17 &   9.0  &    0.48 &     0.14 & Delfosse et al. (2013)\\
 GJ 849b   &  10.4 &   539.5  &   17.4  &   31.0  &    5.1 &  0.04 &   8.8  &    0.49 &     0.99 & Butler et al. (2006)\\
 HIP 79431b & 11.3 &   107.1  &   18.7  &    5.7  &    0.3 &  0.29 &  14.4  &    0.49 &     2.10 & Apps et al. (2010)\\
 HIP 57050b & 11.9 &    13.0  &   18.9  &    0.7  &    0.1 &  0.31 &  11.0  &    0.34 &     0.30 & Haghihipour et al. (2010)\\
 Gl 179b   &  11.9 &   450.0  &   17.5  &   25.7  &    6.3 &  0.21 &  12.3  &    0.36 &     0.82 & Howard et al. (2010)\\
 GJ 317b   &  12.0 &   326.4  &   16.9  &   19.3  &    1.9 &  0.11 &  15.1  &    0.42 &     1.80 & Johnson et al. (2007)\\
\end{tabular}
\label{tab1}
\end{table*}

\subsection{Measuring Transiting Systems Configurations}

\begin{figure}
\centering
\includegraphics[width=0.50\textwidth]{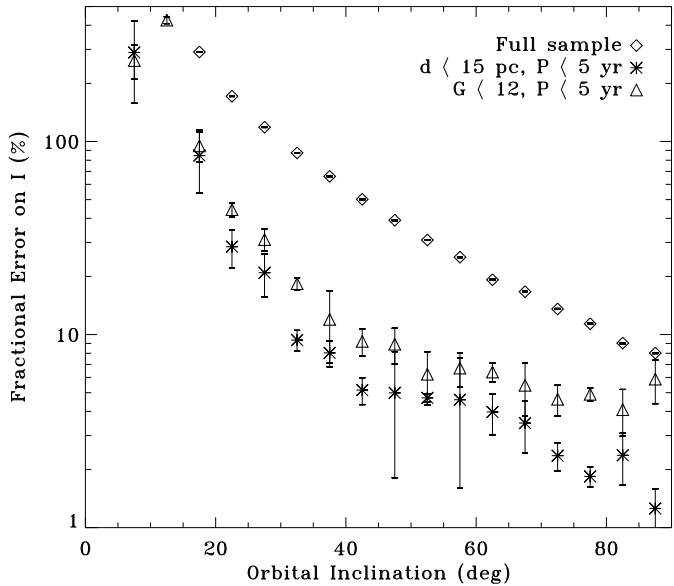} 
 \caption{Fractional error on the inclination angle $i$ as a function of $i$ itself. Error bars correspond to the  standard deviation in each inclination bin. 
 }
\label{fig8}
\end{figure}

The class of transiting planets is of particular importance, as the simultaneous determination of their masses (via Doppler measurements) 
and radii (via transit photometry) provides the means to estimate their densities, a fundamental proxy for understanding their interior compositions 
(e.g., Charbonneau et al. 2007, and references therein). 
Furthermore, if the primaries are sufficiently bright, transiting planets can be further characterized using the 
techniques of transmission and occultation spectroscopy to determine the chemistry and dynamics of their atmospheres (e.g., Seager \& Deming 2010).

 On the one hand, detection of planetary transits is normally achieved via investigation of photometric lightcurves. The general prospects for transiting 
short-period (giant) planet detection with Gaia using its onboard photometry have recently been revisited by Dzigan \& Zucker (2012). On the other hand, 
Gaia high-precision astrometry, by measuring directly the inclination angle 
of an orbit (unlike Doppler spectroscopy), can in principle allow to uncover the existence of a possibly transiting planet 
at wider orbital separations (typically $a>0.5$ AU). We focus here on gauging the potential of Gaia to identify astrometrically extrasolar planets in orbits 
compatible with transit configurations.

We show in Fig.~\ref{fig8} the fractional error in $i$ as a function of $i$ itself as determined in the simulations. The three cases correspond to the 
full  LSPM sub-sample within 33 pc from the Sun, stars within 15 pc and with planets with $P<T$, and stars with $G<12$ and with planets with $P<T$ 
(i.e., using the same selection criteria of \S~\ref{orbit} and the additional constraint of well-sampled orbits). 
The overall trend confirms the findings of Sozzetti et al. (2001), with the Gaia astrometric observations becoming less sensitive to the inclination 
itself as we move towards a quasi-face-on configuration ($i\approx0$ deg), and a corresponding increase of the fractional error on this parameter. 
The immediate conclusion is that $i\simeq90$ deg could be determined with uncertainties of just a few degrees for Jupiter-mass companions on well-sampled orbits 
with $P<T$ around the nearest or brightest M dwarfs. Wider-separation ($a>0.3$ AU) systems with close to edge-on configurations, indicating the presence of a planet 
that might transit and/or be occulted by its primary, would then become very interesting targets for follow-up photometry, to ascertain whether the 
prediction is verified or not. 

 The possibility to study a sample of transiting cold (i.e., long-period) giant planets around nearby low-mass stars is certainly intriguing, for systematic 
comparison with their strongly irradiated, short-period counterparts. While their typical transit depths (significantly exceeding 0.01 mag) would not pose a challenge even for 
modest-precision photometric systems, ground-based transit searches lack sufficient sensitivity at long periods due to the impossibility to guarantee continuous coverage 
over extended time baselines, a necessary prerequisite given that the infrequent transits make it difficult to build enough signal-to-noise ratio. Space-borne instruments can 
fulfill the requirement of uninterrupted photometric coverage, and indeed Kepler has identified transiting giant planet candidates on $\sim1$ AU orbits 
(Fressin et al. 2013). However, these orbit F-G-K dwarfs at hundreds of pc from the Sun, with typical infrared magnitudes of $J\approx12-13$ mag. 
If any such objects were detected around low-mass stars in the solar neighborhood (tens of pc), as they would orbit much brighter primaries at infrared wavelengths, 
they would then constitute prime targets for atmospheric characterization via transit and occultation spectroscopy with future ground-based and particularly 
space-borne instrumentation. 

\begin{figure*}
\centering
$\begin{array}{cc}
\includegraphics[width=0.48\textwidth]{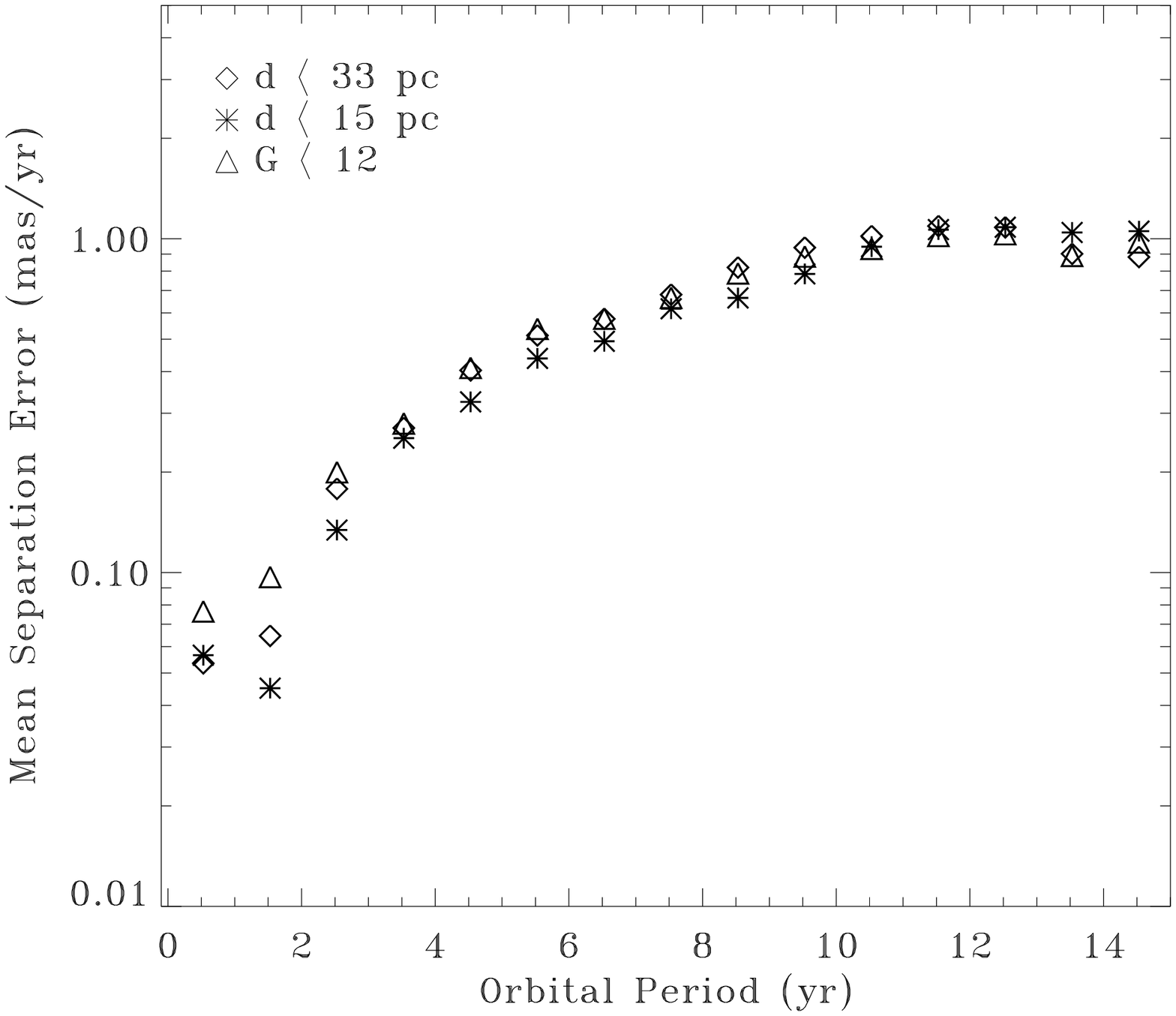} 
\includegraphics[width=0.45\textwidth]{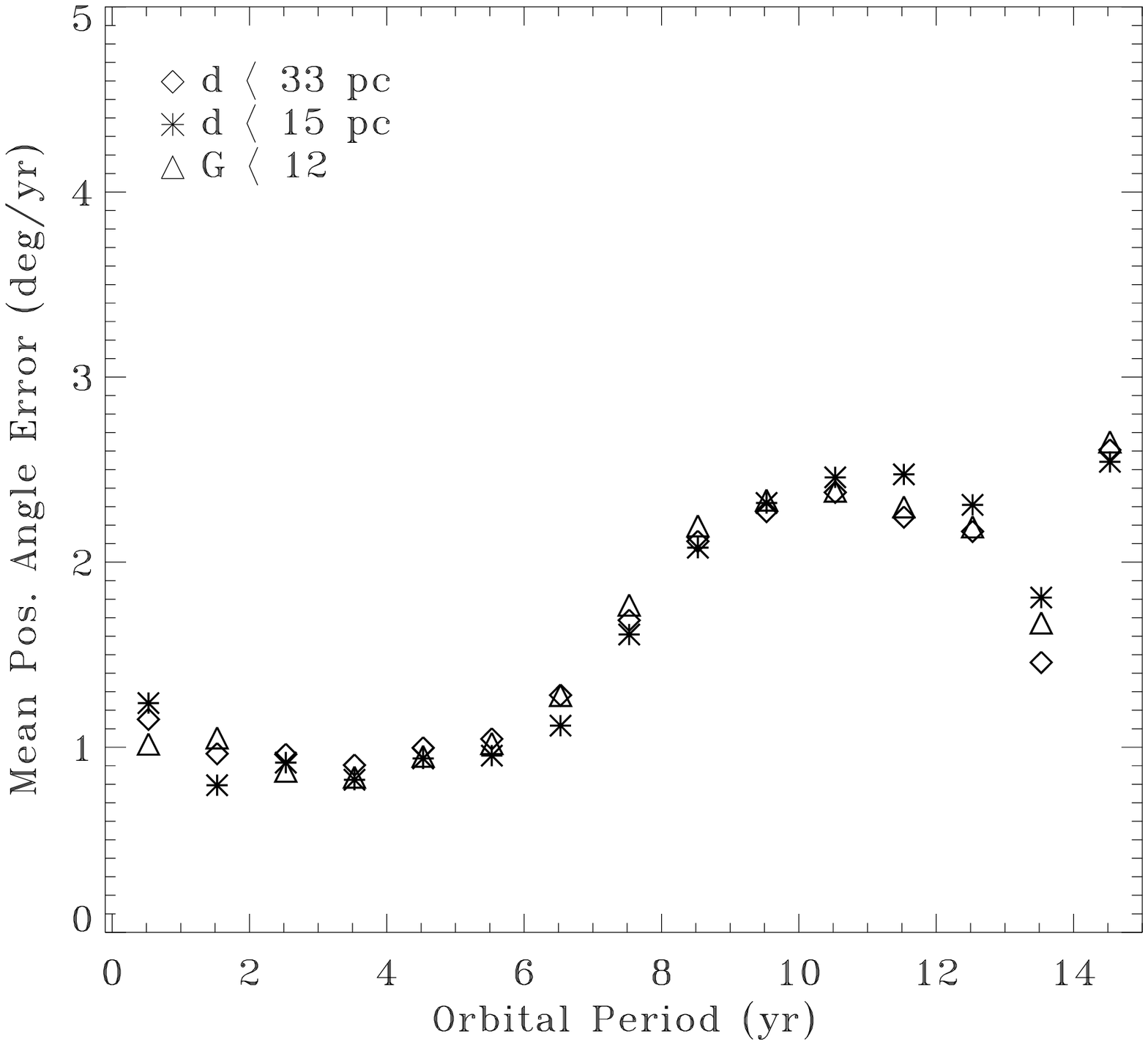} \\
\end{array} $
 \caption{Left: Degradation rate in the prediction for the value of orbital separation of the planet as a function of its orbital period. 
 The results are averages over each 0.5 yr period bin.
 Right: The same, but for the position angle. In both panels, symbols are as in Figure~\ref{fig8}.}
\label{fig9}
\end{figure*}

 For the planet's disc to occult the stellar disc, the orbital inclination must satisfy: 

\begin{equation}
a_p\cos i \leq R_\star + R_p,
\end{equation}

with $R_\star$ and $R_p$ the stellar and planetary radii, respectively. In general, the geometric transit probability can be expressed as (e.g., Barnes 2007): 

\begin{equation}
p_{tr} = 0.0045\left(\frac{1\,AU}{a_p}\right)\left(\frac{R_\star+R_p}{R_\odot}\right)\left[\frac{1+e\cos(\pi/2-\omega)}{1-e^2}\right]
\end{equation}

This expression allows to appreciate how favorable configurations of eccentricity and argument of pericenter can dramatically increase the transit likelihood 
for long-period objects (Kane \& von Braun 2008). However, at large orbital distances (e.g., $a\gtrsim1$ AU) the range of permitted inclination angles for transits to occur 
(given by $i>\arccos(p_{tr})$) is typically well within 1 deg from the central transit configuration ($i=90$ deg), except for pathologically eccentric systems ($e\geq0.9$) 
when $\omega\simeq\pi/2$ (i.e., the pericenter is aligned towards the observer). From Figure~\ref{fig8} we can see that in the case of 
accurately measured orbits (with $\varsigma/\sigma_\mathrm{AL}\gtrsim 10$) a typical fractional uncertainty of $\sim2\%$ on the inclination angle when $i$ is close to 90 deg 
implies formal errors of $\sim2$ deg. That is, a best-case accurate orbit determination by Gaia of a long-period giant planet with $i=90\pm2$ deg would not allow 
to rule out non-transiting configurations, within the errors. At the same time, measuring an astrometric orbit with $i=88\pm2$ deg would also indicate a possibly 
transiting companion, within the errors. Assuming orbits are isotropically distributed, we can convolve this information with the M dwarfs starcounts from the BGM sample 
within 100 pc and the occurrence rate $f_p=3\%$ of giant planets within 3 AU of M dwarfs discussed in \S~\ref{yield} to infer a) the number of 
intermediate-separation giants actually 
transiting their parent stars, and b) the number of giants that might have astrometric orbital solutions compatible with a transiting configuration within $1\sigma$. 

Based on the above considerations, for fractional uncertainties on the inclination angle of 10\%, 5\%, and 2\%, Gaia could detect 255, 85, and 10 systems, respectively, 
formally compatible with transiting configurations within the $1\sigma$ error-bars. On the one hand, using the $f_p=3\%$ estimate the expectation is that 
only 40 systems in the BGM sample would actually have $i$ above the critical value for transits to occur in practice. 
If the sample of actually transiting systems were entirely composed of systems with sufficiently high values of astrometric signals for which Gaia could 
deliver orbits with $i$ determined within 10\% accuracy, then we find that the sample of candidate transiting planets identified by Gaia would encompass $\sim85\%$ 
of false positives. On the one hand, tightening the requirements on the precision with which $i$ can be determined might allow to select a smaller sample of candidates 
with fewer false positives. On the other hand, it might well happen that a candidate system in transit with $i$ measured less precisely is in fact transiting, 
while one with $i$ more accurately measured in fact is not. This will depend in practice on the actual shapes of the period and mass distributions, on the details 
of planet frequency as a function of spectral sub-type, and distance from the Sun of the actual M dwarf sample that will be observed by Gaia. 

Possibly transiting giants planets uncovered astrometrically by Gaia will have to be confirmed by means of follow-up photometric observations, that could readily 
be carried out from the ground even with modest-size telescopes. In perspective, any experiment designed for this purpose will also have to keep the above caveats into 
consideration. For example, such studies would benefit from the availability of additional Doppler measurements aimed at improving the accuracy of the orbital solutions 
and the corresponding transit ephemeris predictions. It will also be important to dentify the correct balance between size of the candidate sample and expectations 
of false positive rates, as such issues could have a significant impact on follow-up programs to verify the actual transiting nature of the detected systems. 
Finally, note that Gaia-detected intermediate-separation giants on orbits compatible with transit configurations might also help revisit the photometric light-curve databases of 
existing (e.g., MEarth, Nutzman \& Charbonneau 2008; APACHE, Giacobbe et al. 2012; Sozzetti et al. 2013) and upcoming (NGTS, Wheatley et al. 2013) ground-based surveys 
 focusing on late-type dwarfs as well as those of other successful programs, such as Super-WASP, HATNet, and HATSouth, looking for missed or uncategorized transit events.

\subsection{Predicting Giant Planets' Location and Brightness}

\begin{figure*}
\centering
$\begin{array}{cc}
\includegraphics[width=0.48\textwidth]{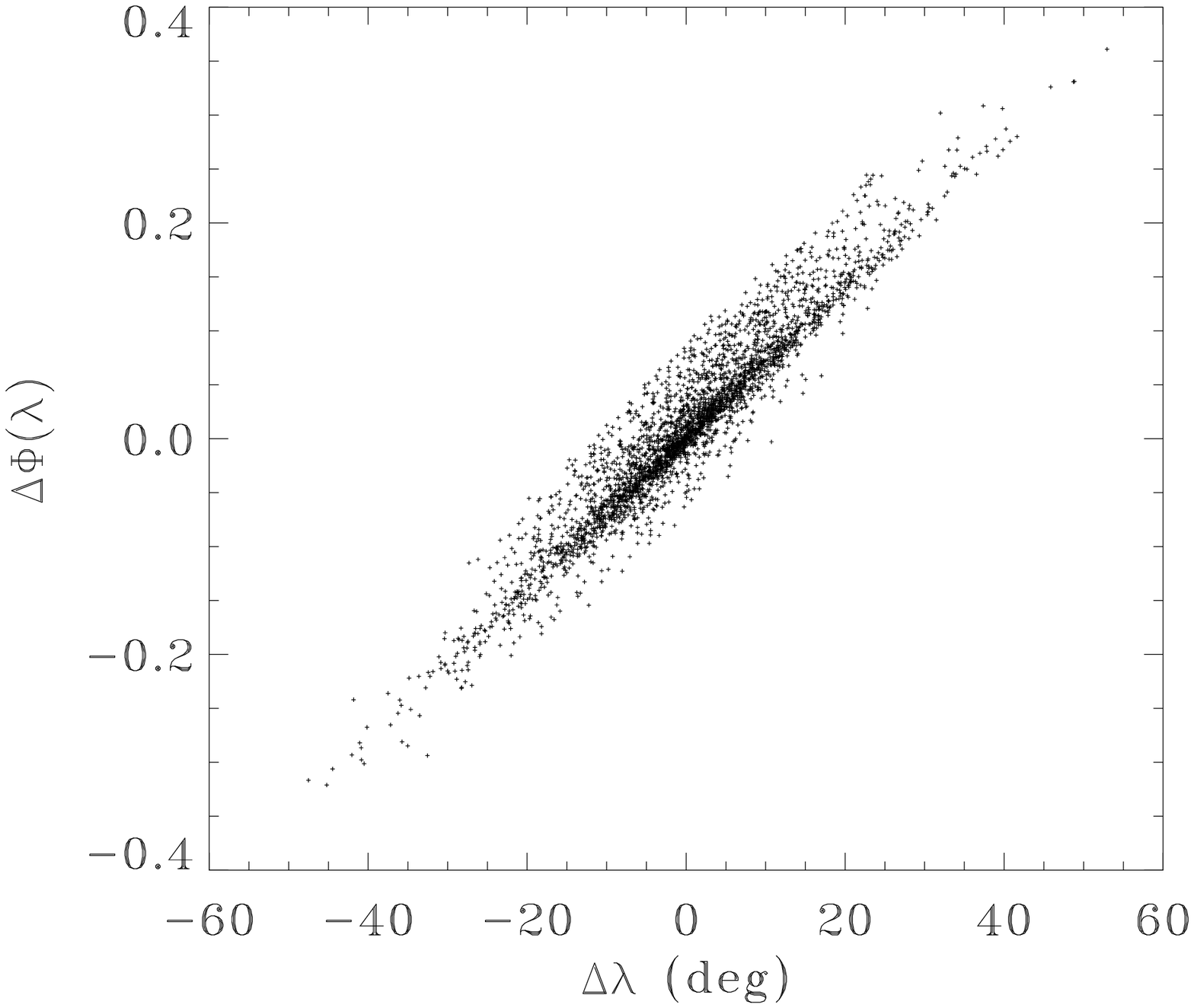} & \includegraphics[width=0.48\textwidth]{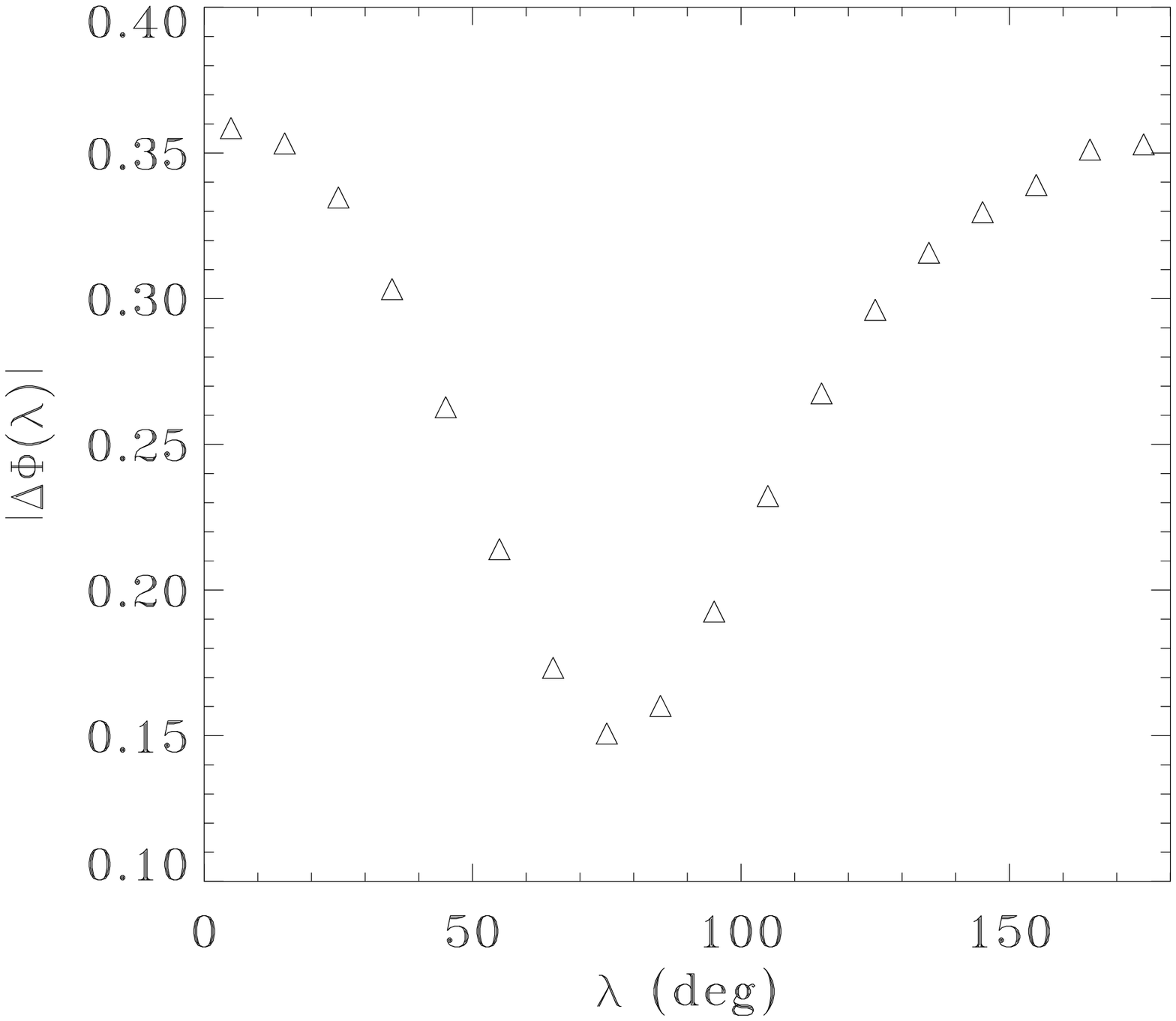} \\
\end{array} $
 \caption{ Left: Errors on the planetary phase function (assuming a Lambert sphere) as a function of the errors on the reconstructed 
 planetary phase angle for the LSPM sub-sample. Each dot represents one target star, and the values correspond to differences taken between orbit-averaged 
 derived and true quantities. 
 Right: Absolute errors $|\Delta\Phi(\lambda)|$ as a function of $\lambda$. The values correspond to the median of all orbital configurations in 10-deg bins in $\lambda$. }
\label{fig10}
\end{figure*}

 As shown by Benedict et al. (2006) using HST/FGS measurements of the $\varepsilon$ Eridani system, astrometry, 
by determining the full orbital geometry and mass of a planetary companion, can have significant value for future direct-imaging programs 
and for the interpretation of emergent flux measurements. 
For example, by determining the times, angular separation and position angle at periastron and apoastron passage, it will be 
possible to predict {\it where} and {\it when} a planet will be at its brightest (this is  also relevant for eccentric planets which 
can undergo orders of magnitude of variation in apparent brightness along the orbit), thus a) crucially helping in the optimization 
of direct imaging observations and b) resolving at least in part important model degeneracies in predictions of an 
exoplanet apparent brightness in reflected host star light as functions of orbit geometry, companion mass, system age, 
orbital phase, cloud cover, scattering mechanisms, and degree of polarization (e.g., Sudarsky et al. 2005; Burrows et al. 2004; Madhusudhan \& Burrows 2012).

The first element of the synergy between Gaia astrometry and future direct-imaging projects consists in being able to quantify the accuracy 
with which it will be possible to predict where to look around a given star, based on the companion mass and orbital parameters determination.
We show in Fig.~\ref{fig9} the average rates of degradation $\Delta\varrho$ and $\Delta\vartheta$ in the estimated  orbital separation $\varrho$ and 
position angle $\vartheta$ of the planet (expressed in  mas yr$^{-1}$ and deg yr$^{-1}$, respectively) as a function of the orbital period. On average, the knowledge 
of the planet's  ephemeris will degrade at rates of  $\Delta\varrho < 1$ mas yr$^{-1}$ and $\Delta\vartheta < 2$ deg yr$^{-1}$, for orbits with $P<T$. 
These numbers are over an order of magnitude smaller than the degradation levels attained by present-day ephemerides predictions based on mas-level 
precision HST/FGS astrometry (Benedict et al. 2006).  In the present sample of intermediate-separation giant planets around M dwarfs (see Table~\ref{tab1}), 
one could then conclude that at least two objects,  GJ 832b and  GJ 433c, with typical separations of $0.7^{\prime\prime}$ and $0.4^{\prime\prime}$, 
respectively, represent prime candidates for such an investigation when Gaia data will become available, particularly if combined with existing 
radial-velocity datasets (thus improving the accuracy of the ephemeris predictions). In this regime of orbital separations instruments such as 
SPHERE on the VLT (Kasper et al. 2012) and particularly PCS on the E-ELT (see https://www.eso.org/sci/facilities/eelt/instrumentation/) are in fact 
expected to achieve very high contrast ratios. 

A second element of the above mentioned synergy relates to the effectiveness with which a precise knowledge of the companion mass and 
orbital geometry of the system from astrometry can be used to predict accurate times of optimal visibility for direct imaging and 
eventually help discriminate between different atmospheric compositions.
Using as an illustrative example that of an isotropically, perfectly reflecting Lambertian surface (e.g., Burrows 2005; Madhusudhan \& Burrows 2012), 
the left panel of Figure~\ref{fig10} shows how the error in the planetary phase function $\Phi(\lambda)$ varies as a function of the error 
in the derived phase angle $\lambda$ based on the orbital parameters determined for each of the 1-M$_J$ companions  around the LSPM M-dwarf sub-sample. 
In the Figure, for each target in the  LSPM sub-sample the errors $\Delta\lambda$ and $\Delta\Phi(\lambda)$ are defined as the difference between the 
orbit-averaged derived value and the orbit-averaged true value of each quantity. For both $\lambda$ and $\Phi(\lambda)$ the median differences are 
very close to zero, indicating that there is little bias in both estimates. Based on the distribution of the absolute differences for the two 
quantities, the median uncertainty on the phase results to be $\Delta\lambda\sim7$ deg, and that on the phase function $\Delta\Phi(\lambda)\sim0.05$.

The rms uncertainty on the phase-averaged phase function is representative of the quality with which this quantity could be determined based on Gaia astrometry 
(assuming a specific atmospheric model), but not at all phases. The right panel of Figure~\ref{fig10} shows how the error in the phase function varies with $\lambda$. 
In the plot, each point corresponds to the median of the absolute differences between the derived and the true value of $\Phi(\lambda)$ in each 10-deg bin in $\lambda$ 
(considering only the subsets of the LSPM sample with planets whose orbits have been sampled in any given interval in phase angle). 
Note that, depending e.g., on orbit geometry and details on the atmospheric scattering properties the value of $\Phi(\lambda)$ 
can vary by a factor of 2 or more at a given phase angle (Sudarsky et al. 2005; Madhusudhan \& Burrows 2012) for intermediate-separation giant planets. 
Based on the result shown in the right panel of Figure~\ref{fig10}, then for a value of the phase angle of say $\lambda=70$ deg, it would be possible to distinguish, 
on average, between the phase function of a Lambert sphere ($\Phi(\lambda)\sim0.5$) and that of Jupiter ($\Phi(\lambda)\sim0.2$) at the $2-\sigma$ level 
(see e.g. Figure 3 of Sudarsky et al. 2005). However, a difference between $\Phi(\lambda)\sim0.9$ for a Lambert sphere and $\Phi(\lambda)\sim0.6$ for Jupiter 
at $\lambda=30$ deg might still fall within the typical errors. 

\begin{figure}
\centering
\includegraphics[width=0.50\textwidth]{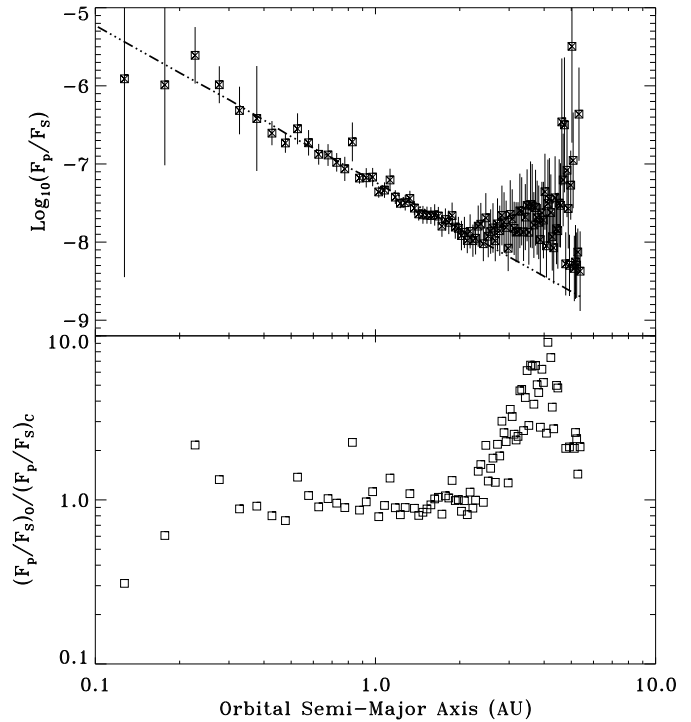} 
 \caption{ Top: Predicted planetary emergent flux as a function of orbital semi-major axis from the M dwarf primary. Logarithmic error bars on $F_p/F_\star$ 
 are phase-averaged values in each 0.05 AU bin obtained through error propagation of the uncertainties on $a_p$ and $\Phi(\lambda)$, assuming an error on 
 the planetary radius ($R_p=1$ $R_J$) of 5\%. Theoretical values based on a Lambert sphere model are shown as a dashed-dotted line. 
 Bottom: The ratio of $(F_p/F_\star)_O$, derived from the simulated data, to $(F_p/F_\star)_C$, computed from the model, as a function of orbital semi-major axis. }
\label{fig11}
\end{figure}

Based on the values of $\Phi(\lambda)$ so determined, we show in the  top panel of Figure~\ref{fig11} the predicted phase-averaged planetary emergent 
flux in units of the stellar flux $F_p/F_\star$ 
as a function of orbital separation from the M dwarf primary.  The simulated data are binned in 0.05 AU intervals, while the theoretical values are 
shown as a dashed-dotted line. The bottom panel of Figure~\ref{fig11} shows the ratio $(F_p/F_\star)_O/(F_p/F_\star)_C$ 
between the simulation results and the Lambert sphere model. Note the close agreement for well-sampled orbits ($a\lesssim2$ AU). 
In this regime of orbital separations, good quality orbit reconstruction implies that $(F_p/F_\star)_O$ is on average within a factor 1.2 of $(F_p/F_\star)_C$. 
At $a\sim3$ AU, $(F_p/F_\star)_O/(F_p/F_\star)_C\approx2$, on average. 
It should be kept in mind that, as shown by e.g. Burrows (2005) and Sudarsky et al. (2005), planet/star flux ratios depend strongly on the 
Keplerian elements of the orbit, particularly inclination and eccentricity, with variations in $F_p/F_\star$ of up to over an order of magnitude. 
Thus for example (assuming the object is detectable by a direct-imaging device) a flux difference of a factor $\sim16$ along the orbit of a 
Jupiter-mass companion on an 1.5-AU, high eccentricity ($e=0.6$) orbit could be inferred with very high significance. Similarly, a flux difference 
of a factor $\sim4$ between a cloud-free giant planet at 2 AU and one with a large-size (100 $\mu$m) condensate particles atmosphere could also be 
detected (see e.g. Figure 16 and 17 of Sudarsky et al. 2005). 
At short orbital distances ($a\lesssim0.3$ AU), the combination of insufficient orbit sampling and low values of $\varsigma/\sigma_\mathrm{AL}$ 
worsens the quality of the determination of $F_p/F_\star$. For $a\gtrsim3$ AU, $F_p/F_\star$ is systematically overestimated  (by up to a factor of ten, and more) 
due to the systematic underestimation of the orbital period.  Finally, we note from Figure~\ref{fig11} that `cold' giant planets on 2-AU orbits are expected 
to have $F_p/F_\star\sim10^{-8}$. Taking the numbers at face value, 
these systems thus appear difficult for an instrument such as SPHERE, nominally set to achieve contrast ratios in the range $10^{-6}-10^{-7}$ (Kasper et al. 2012). 
The combination of Gaia astrometry for such systems might then become more effective with next-generation direct imaging devices on telescopes such as the E-ELT, for 
which contrast ratios on the order of $10^{-8}-10^{-9}$ could be more readily achieved (e.g., Kasper et al. 2010).

\section{Summary and Conclusions}\label{summ}

In this work we report results from a detailed numerical experiment designed to assess the potential of ESA's Cornerstone mission Gaia to 
detect and characterize astrometrically giant planetary companions to our closest neighbors, the reservoir of cool low-mass M dwarfs within $\sim30$ pc from the Sun. 
The paper was motivated by the need to revisit and update Gaia's planet detection potential now that we are within  a few months from launch, relaxing 
some of the caveats and simplifying assumptions of previous analyses (e.g., using an up-to-date Gaia error model and employing an actual list of stars in input). 
A second aim of this work was to begin shedding light on some of the potentially relevant synergies between Gaia astrometry and other ongoing and planned 
planet detection and characterization programs, both from the ground and in space. The results obtained in this work have been specifically tailored 
to a sample of nearby, low-mass M dwarfs. The main findings in this experiment can be summarized as follows: 

\begin{itemize}
\item[1)] the overall quality of orbit reconstruction (using $P$, $M_p$, and $\varsigma/\sigma_\mathrm{AL}$ as proxies) 
is in agreement with previous findings by e.g. Lattanzi et al. (2000), Sozzetti et al. (2002) and Casertano et al. (2008). In particular, the impact of the gate 
scheme to avoid saturation on bright ($G<12$ mag) stars (that tends to worsen the positional accuracy of Gaia in this magnitude range) is found to be relatively mild for 
the M-dwarf  LSPM sub-sample studied  (with an average $G\simeq14$ mag). Given that the current occurrence  rate estimates for giant planets orbiting within 3 AU of low-mass stars are on the order of 3-4\%, 
we can expected $\approx10^2$ massive planetary companions to be detected by Gaia around this sample (a ten-fold increase with respect to present-day giant exoplanet counts). 
Extrapolations based on M dwarfs starcounts out to 100 pc from the Sun (a reservoir of $\sim4\times10^5$ stars) allow us to infer the possibility for Gaia 
to detect over two thousand new giant planets around low-mass stars, and to derive accurate masses and orbital parameters for as many as five hundred systems.
 The size of the sample will likely help to constrain $f_p$ to a very accurate degree: based on the simulations presented here, we estimate that, assuming 
present-day estimates of the occurrence rates, Gaia astrometry would allow to determine $f_p=3\%$ with an error of 2\%, an improvement by over an order of magnitude 
with respect to the most precise estimates to-date. This in turn will allow for meaningful comparisons with theoretical predictions 
of giant planet frequencies in the low stellar mass regime. Such systems will also become prime targets for high-precision RV follow-up from the ground, looking for 
the presence of additional, low-mass companions with state-of-the-art facilities both in the visible (e.g., HARPS, HARPS-N) and in the near infrared (e.g., CARMENES, HPF); 

\item[2)] for detected giant planets with periods in the range $0.2-5$ yr (i.e., with 
accurately determined masses and orbits), inclination angles corresponding to quasi-edge-on configurations will be determined with enough precision (a few percent) 
so that it will be possible to identify candidate transiting planets in a regime of orbital separations which is inaccessible from the ground and only marginally 
probed from space by dedicated transit discovery missions such as CoRoT and Kepler.  Based on the BGM sample results, Gaia might be able to measure accurately the 
orbits of 10 potentially transiting intermediate-separation giants around nearby M dwarfs. Considering inclination angles determined with 10\% accuracy, the sample of 
`astrometric' candidate long-period transiting planets might encompass more than 250 systems. However, the majority of these candidates ($\sim85\%$) would be likely 
false positives. Ground-based monitoring campaigns will be instrumental in unveiling the true nature of the systems.; 

\item[3)] for well-sampled orbits ($P<T$), the uncertainties on planetary ephemerides, 
separation $\varrho$ and position angle $\vartheta$, will degrade at typical rates of  $\Delta\varrho < 1$ mas yr$^{-1}$ and $\Delta\vartheta < 2$ deg yr$^{-1}$, 
respectively. These are over an order of magnitude smaller than the degradation levels attained by present-day ephemerides predictions based on mas-level precision astrometry; 

\item[4)] Planetary phases will be measured with typical uncertainties $\Delta\lambda$ of several degrees, resulting (under the assumption of simple purely 
scattering atmospheres) in phase-averaged errors on the phase function $\Delta\Phi(\lambda)\approx0.05$, and expected  phase-averaged 
uncertainties in the determination of the emergent flux of well-measured, intermediate-separation 
($0.3<a<2.0$ AU) giant planets of  $\sim20\%$. The combination of detailed models of giant exoplanets' systems and reliable ephemerides 
from Gaia astrometry could then greatly help both in the selection of good targets for direct-imaging instruments and for the physical interpretation of positive 
observational results. 

\end{itemize}

Our findings constitute a first step in the characterization of the full impact of Gaia astrometry in the realm of exoplanets orbiting low-mass stars. Indeed, several 
important issues will be worthy of future investigations, particularly now that the launch of Gaia is looming very close. For example, it would be valuable to provide 
an assessment of the effectiveness of the combination of Gaia data with high-precision RVs for the sample of objects listed in Table~\ref{tab1}, and an extension of this 
study to multiple-systems configurations (such as the GJ 876 system) also ought to be carried out. We have assumed all stars in the L\'epine (2005)  LSPM sub-sample used here to be 
single, but this is not likely to be a realistic approximation, and the problem of astrometric planet detection in the presence of orbital motion induced by a distant 
companion star (e.g., Sozzetti 2005) will have to be tackled eventually. It might also be worthwhile to investigate to which extent accurate orbital solutions 
indicating potentially transiting intermediate-separation giant planets could allow to infer precise transit times for successive photometric follow-up. The fine details of the 
synergy resulting by the actual combination of Gaia and direct imaging devices data for improving the interpretation of observables in reflected light (phase curves, 
geometric albedos, polarization parameters) of extrasolar planets in terms of the underlying scattering mechanisms and in turn chemical and 
thermal properties of their atmospheres have also been left largely unexplored. 
Nevertheless, the results presented here help to quantify the actual relevance of the Gaia observations of the large sample of nearby M dwarfs in a synergetic effort 
to optimize the planning and interpretation of follow-up/characterization measurements of the discovered systems by means of transit photometry, and 
upcoming and planned ground-based as well as space-borne observatories for direct imaging (e.g., VLT/SPHERE, E-ELT/PCS) and simultaneous multi-wavelength spectroscopy 
(e.g., EChO, JWST).

\section*{Acknowledgments}

We thank U. Abbas, D. Busonero, and A. Spagna for helpful discussions. This research has made use of the VizieR catalogue access tool, CDS, Strasbourg, France, and 
of NASA's Astrophysics Data System. We gratefully acknowledge partial support from the European Science Foundation (ESF) within the
 'Gaia Research for European Astronomy Training' Research Network Programme. This work has been funded in part by ASI under contract 
 to INAF I/058/10/0 (Gaia Mission - The Italian Participation to DPAC). An anonymous referee provided a thorough, critical review, and 
 very valuable comments and suggestions that significantly improved an earlier version of the manuscript. 

\label{lastpage}

\end{document}